\documentclass[preprint,10pt]{aastex}%
\usepackage{emulateapj5}
\usepackage{epsfig}
\usepackage{amsmath}
\usepackage{amsfonts}
\usepackage{amssymb}
\usepackage{graphicx}%
\def\ra{{\rm R.A. }}
\def\dec{{\rm Dec.}}
\def\deg{^\circ}

\def\aitem#1{\vbox{\hskip0.0in\hbox to 3.5in{\parbox{3in}{---\:{\bf #1}\hfil}}}\vskip0.1in}

\begin{document}

\title{Detection of Polarization in the Cosmic Microwave Background using DASI}

\author{J.\ Kovac, E.\ M.\ Leitch, C.\ Pryke and J.\ E.\ Carlstrom}
\affil{University of Chicago,
Department of Astronomy \& Astrophysics,
Department of Physics,
Enrico Fermi Institute,
5640 South Ellis Avenue,
Chicago, IL 60637}

\smallskip
\author {and}

\author {N.\ W.\ Halverson and W.\ L.\ Holzapfel}
\affil{University of California,
Department of Physics, Le Conte Hall,
Berkeley, CA 94720}

\begin{abstract}
We report the detection of polarized anisotropy in the Cosmic
Microwave Background radiation with the Degree Angular Scale
Interferometer (DASI), located at the Amundsen-Scott South Pole
research station.  Observations in all four Stokes parameters were
obtained within two $3\fdg4$ FWHM fields separated by one hour in Right  
Ascension.  The fields were selected from the subset of fields observed with
DASI in 2000 in which no point sources were detected and are located in 
regions of low Galactic synchrotron and dust emission.
The temperature angular power spectrum is consistent with previous
measurements and its measured frequency spectral index is $-0.01$ ($-0.16$ -- 0.14
at 68\% confidence), where
0 corresponds to a 2.73~K Planck spectrum.  The power spectrum of
the detected polarization is consistent with theoretical predictions
based on the interpretation of CMB anisotropy as arising from primordial
scalar adiabatic fluctuations.  Specifically, $E$-mode polarization is
detected at high confidence ($4.9\sigma$). Assuming a shape for the
power spectrum consistent
with previous temperature measurements, the level found for the
$E$-mode polarization is 0.80 (0.56 -- 1.10), where the 
predicted level given previous temperature data is 0.9 -- 1.1.
At 95\% confidence, an upper limit of 0.59 is set to the level of $B$-mode polarization with the
same shape and normalization as the $E$-mode spectrum.  The $TE$
correlation of the temperature and $E$-mode polarization is detected at 
95\% confidence, and also found to be consistent with
predictions. These results provide strong validation of the underlying
theoretical framework for the origin of CMB anisotropy and lend
confidence to the values of the cosmological parameters that have been
derived from CMB measurements.
\end{abstract}

\section{Introduction}

Measurements of the Cosmic Microwave Background (CMB) radiation reveal
the conditions of the universe when it was $\sim400,000$ years old 
with remarkable precision. 
The three most fundamental properties of the CMB are
its frequency spectrum and the angular power spectra of
the temperature and polarization fluctuations.  The frequency
spectrum was well determined by the COBE FIRAS instrument \markcite{mather94,fixsen96}({Mather} {et~al.} 1994; {Fixsen} {et~al.} 1996). The
initial detection of temperature anisotropy was made on large
angular scales by the COBE DMR instrument \markcite{smoot92}(Smoot {et~al.} 1992) and recently there
has been considerable progress in measuring the anisotropy on finer angular
scales
\markcite{miller99,halverson02,netterfield02,lee02}({Miller} {et~al.} 1999; {Halverson} {et~al.} 2002; {Netterfield} {et~al.} 2002; {Lee} {et~al.} 2001).
There have been many efforts to measure the polarization (see below)
but so far, detection of this property of the
CMB has remained beyond the reach of the most sensitive observations.

In the past several years, a standard cosmological model has 
emerged \markcite{hudodelson02}(see, e.g., {Hu} \& {Dodelson} 2002).
In this model, the structure of the CMB angular power spectrum at
degree angular scales is assumed to arise from acoustic oscillations of
the photon-baryon fluid sourced by primordial scalar adiabatic
fluctuations. At decoupling, the modes at maximal amplitude lead to excess
power in the observed CMB angular power spectrum resulting in a harmonic
series of peaks and troughs.  
Within this theoretical framework, and given knowledge of the
temperature angular power spectrum, a prediction can be made for
the level of the CMB polarization with
essentially no free parameters
\markcite{kaiser83,bond84,polnarev85,kamionkowski97,zaldarriaga97,hu_w97,kosowsky99}({Kaiser} 1983; Bond \& Efstathiou 1984; {Polnarev} 1985; Kamionkowski, Kosowsky, \&  Stebbins 1997; {Zaldarriaga} \& {Seljak} 1997; Hu \& White 1997; {Kosowsky} 1999). 
A determination of the CMB polarization would therefore provide a
critical test of the underlying theoretical framework
\markcite{huspergelwhite97,kinney01,bucher01}({Hu}, {Spergel}, \& {White} 1997; {Kinney} 2001; {Bucher}, {Moodley}, \& {Turok} 2001) and therefore of the validity of
cosmological parameters derived from CMB measurements.
Polarization measurements also offer the potential to
triple the number of observed CMB quantities and to enhance our
ability to constrain 
cosmological parameters.

CMB polarization arises from Thompson scattering by electrons of a
radiation field with a local quadrupole moment \markcite{rees68}({Rees} 1968). 
In the primordial plasma, 
the local quadrupole moment is suppressed until the photon mean free
path grows during decoupling. At this time, the largest contribution to the
local quadrupole is due to 
Doppler shifts induced by the velocity field of the plasma \markcite{zaldarriaga_h95}(Zaldarriaga \& Harari 1995). 
In this way, CMB polarization directly
probes the dynamics at the epoch of decoupling. For a Fourier mode of the
acoustic oscillations, the electron velocities are perpendicular to
the wavefronts, leading to either a parallel or perpendicular alignment of 
the resulting polarization.
These polarization modes are referred
to as the scalar $E$-modes in analogy with electric fields; they have no curl component.
Since the level of the polarization depends on velocity, one expects
that the peaks in the scalar $E$-mode power
spectrum correspond to density modes that are at their
highest velocities at decoupling and are therefore at minimum amplitude.
The location of the harmonic peaks in the
scalar $E$-mode power spectrum are therefore expected to be 
out of phase with the peaks in the temperature spectrum
\markcite{kamionkowski97,zaldarriaga97,hu_w97}(Kamionkowski {et~al.} 1997; {Zaldarriaga} \& {Seljak} 1997; Hu \& White 1997).

Primordial gravity waves will lead to polarization in the 
CMB \markcite{polnarev85,crittenden93}({Polnarev} 1985; {Crittenden}, {Davis}, \&  {Steinhardt} 1993) with an
$E$-mode pattern as for the scalar density perturbations, but will also 
lead to a curl
component, referred to as $B$-mode polarization
\markcite{kamionkowski97,seljak97}(Kamionkowski {et~al.} 1997; {Seljak} \& {Zaldarriaga} 1997).  The $B$-mode component is due to
the intrinsic polarization of the gravity waves. 
In inflationary models, the level of the gravity wave induced
$B$-mode polarization power is set by the energy scale of inflation 
to the fourth power. 
While the detection of $B$-mode polarization would provide a critical
test of inflation, the signal is likely to be very weak and may
have an amplitude that is effectively unobservable \markcite{lyth97}({Lyth} 1997). 
Furthermore, distortions to the
scalar $E$-mode polarization by the gravitational lensing of the intervening large
scale structure in the universe will lead to a contaminating B-mode
polarization signal which will severely complicate the extraction of the
gravity-wave induced signal \markcite{zaldarriaga98,hu02,knox02}({Zaldarriaga} \& {Seljak} 1998; {Hu} \& {Okamoto} 2002; {Knox} \& {Song} 2002). 
The possibility, however, of
directly probing the universe at energy scales of $\sim10^{16}$ GeV
by measuring the 
gravity-wave 
induced polarization \markcite{kamionkowski99}(see, e.g., Kamionkowski \& Kosowsky 1999)
is a compelling goal for CMB polarization observations.

Prior to the results presented in this paper, only upper limits have
been placed on the level of CMB polarization. This
is due to the low level of the expected signal, demanding
sensitive instruments and careful attention to 
sources of systematic uncertainty \markcite{staggs99}(see {Staggs}, {Gunderson}, \& {Church} 1999, for a review of CMB polarization measurements).

The first limit to the degree of polarization of the CMB was set by
\markcite{penzias65}{Penzias} \& {Wilson} (1965) who stated that the new radiation that they had discovered was
isotropic and unpolarized within the limits of their
observations. Over the following 20 years, dedicated polarimeters have been
used to set much more stringent upper limits on angular scales of order 
several degrees and larger
\markcite{caderni78,nanos79,lubin79,lubin81,lubin83,sironi97}({Caderni} {et~al.} 1978; {Nanos} 1979; {Lubin} \& {Smoot} 1979, 1981; {Lubin}, {Melese}, \& {Smoot} 1983; Sironi {et~al.} 1997). 
The current best upper limits for the $E$-mode and  $B$-mode
polarizations on large angular scales are 10$~\mu$K at 95\% confidence
for the multipole range $2 \le l \le 20$, set by the POLAR
experiment \markcite{keating01}({Keating} {et~al.} 2001). 

On angular scales of order one degree, \markcite{wollack93}{Wollack} {et~al.} (1993) used the
Saskatoon experiment to set the first upper limit to the CMB
polarization ($25~\mu$K at 95\% confidence for $l \sim 75$); this
limit is also noteworthy in that it was the first limit that was lower
than the level of the CMB temperature anisotropy. The current best limit on
similar angular scales
was set by the PIQUE experiment
\markcite{hedman02}({Hedman} {et~al.} 2002), who reported a 95\% confidence upper limit of
$8.4~\mu$K to the $E$-mode signal, assuming no B-mode polarization.
\markcite{cartwright02}{Cartwright} {et~al.} (2002) presented a preliminary analysis of
CBI data that indicated
an upper limit similar to the PIQUE result, but on somewhat smaller 
scales.

On much finer angular scales of order an arcminute, polarization
measurements have also been pursued and upper limits set
\markcite{partridge97,subrahmanyan00}(e.g., {Partridge} {et~al.} 1997; {Subrahmanyan} {et~al.} 2000).  However, at these angular scales,
corresponding to multipoles $\sim 5000$, the level of the
primary CMB anisotropy is
strongly damped and secondary effects
due to the interactions with large scale structure in the
universe are expected to dominate \markcite{hudodelson02}({Hu} \& {Dodelson} 2002).

In this paper, we present the results of CMB polarization measurements made
with the Degree Angular Scale Interferometer (DASI) located at the NSF
Amundsen-Scott South Pole research station during the 2001 and 2002 austral winter
seasons. DASI was successfully used to measure the temperature anisotropy
from $140 < l < 900$ during the 2000 
season. Details of the instrument, the measured power spectrum and the
resulting cosmological constraints were presented in a series of three papers
~\markcite{leitch02a,halverson02,pryke02}(Leitch {et~al.} 2002b; {Halverson} {et~al.} 2002; {Pryke} {et~al.} 2002, hereafter Papers, I, II and III, respectively). Prior to the start of the 2001 season,
DASI was modified to allow polarization measurements in all four Stokes
parameters over the same $l$ range as the previous measurements. The modifications to the instrument,
observational strategy, data calibration and data reduction are discussed in
detail in 
\markcite{leitch02b}Leitch {et~al.} (2002a) (hereafter Paper IV).

This paper is organized as follows. In  \S\ref{sec:instrument}
we briefly summarize the modifications to the instrument, the observing strategy
and the data calibration from Paper IV and in \S\ref{sec:observations}
the CMB observations and data reduction are discussed. 
The noise model and detection of signal in our
data are discussed in \S\ref{sec:noiseandsignal}. The analysis
method is presented in  \S\ref{sec:lhformalism} and the  
results from the likelihood analysis which models and parameterizes the
signal in terms of CMB polarization and temperature angular power spectra 
are given in \S\ref{sec:lhresults}. 
In \S\ref{sec:systematics} we discuss systematic uncertainties including
instrumental effects and the possibility of foreground contamination.
Conclusions are summarized in \S\ref{sec:conclusion}.

\section{Measuring Polarization with DASI}

\label{sec:instrument}

DASI is a compact interferometric array optimized for the measurement of CMB 
temperature and polarization anisotropy.
A detailed discussion of the DASI instrument is given
in Paper I and the modifications for polarimetric observations
are given in Paper IV. Here we briefly summarize the aspects
of the instrument particularly relevant for polarization
measurements.

Because they directly sample Fourier components of the sky, interferometers 
are uniquely suited to measurements of the CMB angular power spectrum.
In addition, an interferometer gathers instantaneous two dimensional information
while inherently rejecting large-scale gradients in atmospheric 
emission.
For observations of CMB polarization, interferometers offer several additional 
advantages. 
They can be constructed with the required small and stable instrumental polarization.
Furthermore, linear combinations of the data can be used to construct
quantities with essentially pure $E$- and $B$-mode polarization response patterns
on a variety of scales. 
This property of the data greatly facilitates the analysis and interpretation of 
the observed polarization in the context of cosmological models.

DASI consists of 13 horn antennas mounted on a rigid 
faceplate in a three-fold symmetric pattern. 
Following each antenna is a cooled, low-noise HEMT amplifier 
optimized for the 26 -- 36~GHz DASI band.
The signals from the receivers are combined in a wideband analog correlator 
that computes the complex correlation, or {\it visibilities}, between pairs of receivers in 
ten 1-GHz bands.  
The locations of the horns in the faceplate are optimized to provide
uniform sampling over the multipole range $l\sim $140 -- 900. 
The DASI mount is designed to track in elevation and azimuth with the 
additional ability to rotate the entire horn array about the faceplate axis. 
The flexibility of this mount allows us to tailor the Fourier $(u,v)$ plane coverage of 
a given observation as well as perform sensitive tests for instrumental 
offsets and other possible systematic errors.

As was detailed in Paper I and \markcite{lay00}{Lay} \& {Halverson} (2000), the sky above the South
Pole is extremely dry and stable, resulting in atmospheric contamination
far below DASI's instrumental noise. 
In addition to the
excellent observing conditions at the Pole, it is possible to track a
single field continuously at constant elevation angle.  
These two characteristics of the Pole enable
the deep integrations needed to detect the CMB
polarization signal.  In these sensitive observations, sources of
foreground emission are potentially a serious concern. Fortunately,
the sky above the Pole includes regions with low diffuse
foreground emission.

\subsection{Hardware Upgrade}

During the 2000--2001 austral summer, DASI's thirteen receivers were
retrofit with broadband achromatic circular polarizers.  These polarizers
employ a multi-element design capable of rejecting the unwanted
polarization state to better than $1\%$ across DASI's frequency band
(see Paper IV, also \S\ref{sec:onaxisleak}).

An interferometer measures the correlations between the signals from
pairs of receivers; as indicated by Equation~\ref{eq:vis} in \S \ref{sec:theory}, 
recovery of the
full complement of Stokes parameters requires the correlation of all
four pairwise combinations of left and right circular 
polarization states ($RR$, $LL$, $RL$
and $LR$), which we refer to as {\it Stokes states}.  The co-polar
states ($RR$, $LL$), are sensitive to the total intensity, while
the cross-polar states ($RL$, $LR$)
measure linear
combinations of the Stokes parameters $Q$ and $U$.

Each of DASI's analog correlators can accommodate only a single
Stokes state, so measurement of the four combinations is achieved via
time-multiplexing.  
The polarizers for each receiver are switched on a
1-hour Walsh cycle, with the result that over the full period of the
cycle, every pair of receivers spends an equal amount of time in all four
Stokes states.

\subsection{Gain and Phase Calibration}
\label{sec:gainphase}

In Paper IV, we detail the calibration of the polarized visibilities
for an interferometer.  In order to produce the calibrated visibilities
as defined in Equation~\ref{eq:vis} below, a complex gain factor $G^S$ which depends on the
Stokes state $S$ must be applied to each raw visibility.
Although the cross-polar gain factors could easily be
determined with observations of a bright polarized source, no
suitable sources are known, and we therefore derive the full calibration
through observations of an unpolarized source. 
The gains for a given pair of receivers $m-n$ (a baseline) can be decomposed into
antenna-based terms (for example, $G_{mn}^{RL} = g_m^R{g_n^{L}}^*$), allowing
us to construct the cross-polar gains from $g_m^R$ and $g_n^L$ derived
from the co-polar visibilities.  DASI's calibration is based on daily
observations of the bright HII region RCW38, described at length in
Paper I, from which we can determine baseline gains for all Stokes
states to better than 2\%.

The procedure for deriving the baseline gains from antenna-based terms
leaves the cross-polar visibilities multiplied by an undetermined
overall phase offset (independent of baseline).  This phase
offset effectively mixes $Q$ and $U$, and must be measured to obtain a
clean separation of CMB power into $E$- and $B$-modes.  Calibration of
the phase offset requires a source whose polarization angle is known,
and we create one by observing RCW38 through polarizing wire grids attached to
DASI's thirteen receivers.  From the wire-grid observations, we can
derive the phase offset in each frequency band with an uncertainty of
$\lesssim 0\fdg4$.

As an independent check of this phase offset calibration, the Moon was
observed at several epochs during 2001--2002.  Although the expected
amplitude of the polarized signal from the Moon is not well known at
these frequencies, the polarization pattern is expected to be radial
to high accuracy, and this can be used to determine the cross-polar
phase offset independently of the wire grid observations.  As shown in
Paper IV, these two determinations of the phase offset are in
excellent agreement.

\subsection{Instrumental Polarization}
\label{sec:instrpol}

\subsubsection{On-Axis Leakage}
\label{sec:onaxisleak}

For idealized polarizers, the cross-polar visibilities are strictly
proportional to linear combinations of the Stokes parameters $Q$ and
$U$.  For realistic polarizers, however, imperfect rejection of the
unwanted polarization state leads to additional terms in the
cross-polar visibilities proportional to the total intensity $I$.
These {\it leakage} terms are the sum of the complex leakage of the
two antennas which form each baseline.  Prior to installation on the
telescope, the polarizers were tuned to minimize these
leakages.

At several epochs during 2001--2002, the antenna-based leakages were
determined with a fractional error of $0.3\%$ from deep observations
of the calibrator source RCW38.  As is shown in Paper IV,
antenna-based leakages are $\lesssim 1\%$ (of $I$) at all frequency bands
but the highest, where they approach $2\%$; this performance
is close to the theoretical minimum for this polarizer design.
Comparison of the measurements from three epochs separated by many
months indicates that the leakages are stable with time.

For observations of the CMB, the presence of leakage has the effect of
mixing power from temperature into polarization in the uncorrected
visibilities.  Given the low level of DASI's leakages, this is
expected to be a minor effect at most (see \S \ref{sec:systematics}).
Nonetheless, in the analysis presented in this paper, the cross-polar
data have in all cases been corrected to remove this effect using the
leakages determined from RCW38.

\subsubsection{Off-Axis Leakage}
\label{sec:offaxisleak}

Although the polarizers were optimized for low on-axis leakage
response, the feeds themselves will induce an instrumental
polarization which varies across the primary beam.  Offset
measurements of RCW38 and the Moon indicate that the off-axis
instrumental polarization pattern is radial, rising from zero at the
beam center to a maximum of $\sim0.7\%$ near $3\deg$, and then tapering to zero
(see also Paper IV).

With the on-axis polarizer leakage subtracted to $\lesssim0.3\%$ (see
above), this residual leakage, while still quite small compared to the
expected level of polarized CMB signal (again, see 
\S\ref{sec:systematics}), is the dominant instrumental contribution.
Although the visibilities cannot be individually corrected to remove
this effect (as for the on-axis leakage), it may be incorporated in
the analysis of the CMB data.  Using fits to the offset data (see
Paper IV for details), we account for this effect by modeling the
contribution of the off-axis leakage to the signal covariance matrix
in the likelihood analysis described in \S\ref{sec:lhformalism}.

\section{CMB Observations and Data Reduction}
\label{sec:observations}

\subsection{Observations}

For the observations presented here, two fields separated by one hour
of Right Ascension, at $\ra = 23^h30^m$ and $\ra = 00^h30^m$, $
\dec = -55\deg$, were tracked continuously.  The telescope alternated
between the fields every hour, tracking them over precisely the same
azimuth range so that any terrestrial signal can be removed by
differencing.  Each 24-hour period comprised 20 hours of CMB
observations and $2.3$ hours of bracketing calibrator observations,
with the remaining time spent on skydips and miscellaneous calibration
tasks.

The fields were selected from the 32 fields previously observed by
DASI for the absence of any detectable point sources (see Paper I).
The locations of the 32 fields were originally selected 
to lie at high elevation angle and to coincide with 
low emission in the IRAS 100 micron and 408~MHz maps of the sky
\markcite{haslam81}({Haslam} {et~al.} 1981).

The data presented in this paper were acquired from 2001 April 10 to 2001 October 27, and
again from 2002 February 14 to 2002 July 11.  In all, we obtained 162
days of data in 2001, and 109 in 2002, for a total of 271 days
before the cuts described in the next section.

\subsection{Data Cuts}
\label{sec:cuts}

Observations are excluded from the analysis, or \emph{cut}, if they
are considered suspect due to hardware problems, inadequate
calibration, contamination from Moon or Sun signal, poor weather, or
similar effects.  In \S\ref{sec:noiseandsignal}, we describe
consistency statistics that are much more sensitive to these effects
than are the actual results of our likelihood analysis, allowing us to
be certain that the final results are unaffected by
contamination.  Here we briefly summarize the principal categories of
data cuts; each cut is described in detail in Paper IV.

In the first category of cuts, we reject visibilities for which
monitoring data from the telescope indicate obvious hardware
malfunction, or simply non-ideal conditions.  These include cryogenics
failure, loss of tuning for a receiver, large offsets between
real/imaginary multipliers in the correlators, and mechanical glitches
in the polarizer stepper motors.  All data are rejected for a
correlator when it shows evidence for large offsets, or
excessive noise.  An additional cut, and the only one based on the
individual data values, is a $> 30\sigma$ outlier cut to reject rare
($\ll0.1\%$ of the data) hardware glitches.  Collectively these cuts
reject $\sim26\%$ of the data.

In the next category, data are cut on the phase and amplitude
stability of the calibrator observations.  Naturally, we reject
data for which bracketing calibrator observations have been lost due
to previous cuts.  These cuts reject $\sim5\%$ of the data.

Cuts are also based on the elevation of the Sun and Moon.
Co-polar data are cut whenever the Sun was above the horizon, and
cross-polar data whenever the solar elevation exceeded $5\deg$.  
These cuts reject 8\% of the data.

An additional cut, which is demonstrably sensitive to poor weather, is based on the significance
of data correlations as discussed in \S\ref{sec:noisemodel}.  
An entire day is cut if the maximum correlation exceeds $8\sigma$.
A total of 22 days are cut by this test in addition to
those rejected by the solar and lunar cuts.

\subsection{Reduction}

Data reduction consists of a series of steps to calibrate and reduce the
dataset to a manageable size for the likelihood analysis. Phase and
amplitude calibrations are applied to the data based on bracketing
observations of our primary celestial calibrator, RCW38. The raw 8.4-s
integrations are combined over each 1-hr observation for each of 6240
visibilities (78 complex baselines $\times$ 10 frequency bands
$\times$ 4 Stokes states). Leakage corrections are applied to the
data, and sequential observations of the two fields in the same
$15^\circ$ azimuth range are differenced to remove any common ground
signal. Except in the case of the sum and difference data used for
the $\chi^2$ consistency tests in \S\ref{sec:chi2tests}, observations from
different faceplate rotation angles, epochs, and azimuth ranges are combined,
as well as the two co-polar Stokes states, $LL$ and $RR$.  The
resulting dataset has $N\leq4680$ elements ($6240 \times 3/4 =
4680$).
We call this the uncompressed dataset, and it contains all of the information
in our observations of the differenced fields for Stokes parameters $I$, $Q$, and $U$.

\section{Data Consistency Tests and $\chi^{2}$ Results}

\label{sec:noiseandsignal}

We begin our analysis by arranging the data into a vector, considered
to be the sum of actual sky signal and 
instrumental noise: $\mathbf{\Delta}=\mathbf{s+n.}$  The noise vector
$\mathbf{n}$ is hypothesized to be Gaussian and random, with zero mean,
so that the noise model is completely specified by a known covariance
matrix $\mathbf{C}_{N} \equiv\left\langle\mathbf{nn}^{t}\right\rangle $.  Any significant excess variance
observed in the data vector $\mathbf{\Delta}$ will be interpreted as
signal. In the likelihood analysis of the next section, we 
characterize the total covariance of the dataset
$\mathbf{C=C}_{T}\left( \mathbf{\kappa}\right) +\mathbf{C}_{N}$ in
terms of parameters $\mathbf{\kappa}$ that specify the covariance
$\mathbf{C}_{T}$ of this sky signal.  This is the conventional
approach to CMB data analysis, and it is clear that for it to
succeed, the assumptions about the noise model and the accuracy of the
noise covariance matrix must be thoroughly tested.  This is
especially true for our dataset, in which we have achieved 
unprecedented levels of sensitivity in an attempt to measure the very
small signal covariances expected from the polarization of the CMB.

\subsection{Noise Model}

\label{sec:noisemodel}

The DASI instrument and observing strategy are designed to remove
systematic errors through multiple levels of differencing.
Slow and fast phase switching as well as
field differencing are used to minimize potentially variable systematic
offsets that could otherwise contribute a non-thermal component to the
noise. 
The observing strategy also includes Walsh sequencing of the
Stokes states, observations over multiple azimuth ranges and faceplate
rotation angles, and repeated observations of the same visibilities on the
sky throughout the observing run to allow checks for systematic
offsets and verification that the sky signal is repeatable.  We
hypothesize that after the cuts described in the previous section, the noise
in the remaining data are Gaussian and white, with no noise
correlations between different baselines, frequency bands,
real/imaginary pairs, or Stokes states. We have carefully
tested the noise properties of the data to validate the use of this model.

Noise variance in the combined data vector is estimated by calculating
the variance in the 8.4-s integrations over the period of 1~hr,
before field differencing. To test that this noise estimate is
accurate, we compare three different short timescale noise estimates: 
calculated 
from the 8.4-s integrations over the 1-hr observations before and after field
differencing and from sequential pairs of 8.4-s integrations.
We find that all three agree within 0.06\% for co-polar data and 0.03\% for 
cross-polar data, averaged over all visibilities after data cuts.  

We also compare the noise estimates based on the short timescale noise
to the variance of the 1-hr binned visibilities over the entire
dataset (up to 2700 1-hr observations, over a period spanning 457
days).  The ratio of long timescale
to short timescale noise variance, averaged over all combined
visibilities after data cuts, is $1.003$ for the co-polar data and
$1.005$ for the cross-polar data, remarkably close to unity. Together with
the results of the $\chi^{2}$ consistency tests described in
\S\ref{sec:chi2tests}, these results demonstrate that the noise is white and
integrates down from timescales of a few seconds to thousands of
hours. We find that scaling the diagonal noise by 1\%
makes a negligible difference in the reported likelihood results 
(see \S\ref{sec:systematics}).

To test for potential off-diagonal correlations in the noise, we
calculate a $6240 \times 6240$ 
correlation coefficient matrix from the 8.4-s
integrations for each day of observations. 
To
increase our sensitivity to correlated noise, we use only
data obtained simultaneously for 
a given pair of data vector elements.  Due to the variable
number of 8.4-s integrations $M$ used to calculate each off-diagonal
element, we assess the significance of the correlation
coefficient in units of $\sigma= 1/\sqrt{M-1}$.  Our weather cut
statistic is the daily maximum off-diagonal correlation coefficient
significance (see \S\ref{sec:cuts}).

We use the mean data correlation coefficient matrix over all days,
after weather cuts, to test for significant correlations over the
entire dataset. We find that 1864 (0.016\%) of the  
off-diagonal elements exceed a significance of $5.5\sigma$, when about
one such event is expected for uncorrelated Gaussian noise.  The
outliers are dominated by correlations between real/imaginary pairs of
the same baseline, frequency band, and Stokes state, and between
different frequency bands of the same baseline and Stokes state. For
the real/imaginary pairs, the maximum correlation coefficient
amplitude is 0.14, with an estimated mean amplitude of 0.02; for
interband correlations the maximum amplitude and estimated mean are
0.04 and 0.003, respectively. We have tested the inclusion
of these
correlations in the likelihood analysis and find that they have a
negligible impact on the results, see \S\ref{sec:systematics}.

\subsection{$\chi^2$ tests}

\label{sec:chi2tests}

As a simple and sensitive test of data consistency, we construct a
$\chi^2$ statistic from various splits and subsets of the visibility
data.  Splitting the data into two sets of observations that should
measure the same sky signal, we form the statistic for both the sum
and difference data vectors,
$\chi^{2}=\mathbf{\Delta}^{t}\mathbf{C}_{N}^{-1}\mathbf{\Delta}$,
where $\mathbf{\Delta} = \left( \mathbf{\Delta_1} \pm
\mathbf{\Delta_2} \right) / 2$ is the sum or difference data vector,
and $\mathbf{C}_{N}= \left( \mathbf{C}_{N1}+\mathbf{C}_{N2} \right)
/4$ is the corresponding noise covariance matrix.  We use the
difference data vector, with the common sky signal component
subtracted, to test for systematic offsets and mis-estimates of the
noise.  The sum data vector is used to test for the presence of a sky signal
in a straightforward way that is independent of the analysis method
used for parameter extraction.

We split the data for the difference and sum data vectors in 
five different ways: 
\begin{enumerate}
\item Year -- 2001 data vs.\ 2002 data,
\item Epoch -- the first half of observations of a given visibility vs.\ 
the second half,
\item Azimuth range -- east five vs.\ west five observation azimuth ranges,
\item Faceplate position -- observations at a faceplate rotation angle of 0$^\circ$ vs.\ 
a rotation angle of 60$^\circ$, and
\item Stokes state -- co-polar observations in which both polarizers
are observing left circularly polarized light ($LL$ Stokes state)
vs.\ those in which both are observing right circularly polarized light
($RR$ Stokes state).

\end{enumerate}

These splits were done on the combined 2001/2002 dataset and (except
for the first split type) on 2001 and 2002 datasets separately, to
test for persistent trends or obvious differences between the
years. The faceplate position split is particularly powerful, since the
six-fold symmetry of the $(u,v)$ plane coverage allows us to measure a
sky signal for a given baseline with a different pair of receivers,
different backend hardware, and at a different position on the
faceplate with respect to the ground shields, and is therefore
sensitive to calibration and other offsets that may depend on these
factors. The co-polar split tests the amplitude and phase calibration
between polarizer states, and tests for the presence of circularly
polarized light. 

For each of these splits, different subsets can be examined: co-polar
data only, cross-polar data only (for all except the Stokes state split),
various $l$-ranges (as determined by baseline length in units of
wavelength), and subsets formed from any of these which isolate modes with the
highest expected signal to noise (s/n). 
These high s/n (sub)subsets must assume some theoretical signal template in defining
the s/n eigenmode basis \markcite{bond97}({Bond}, {Jaffe}, \& {Knox} 1998) in which to arrange the data elements
of the original subset, and for this we use the concordance model defined
in \S\ref{sec:lpar}, although we find the results are not strongly dependent 
on choice of model.  Note that the definitions of which modes are included in
the high s/n subsets are made in terms of average theoretical signal, without
any reference to the actual data.
In Table~\ref{tab:chi2}, we present the difference and sum $\chi^2$
values, for a representative selection of splits and subsets. In each case we give
the degrees of freedom, $\chi^2$ value, and probability to exceed
(PTE) this value in the $\chi^2$ cumulative distribution function. For
the 296 different split/subset combinations that were examined, the 
$\chi^2$ values for the difference data appear consistent with noise;
among these 296 difference data $\chi^2$'s, there are two with a PTE~$< 0.01$
(the lowest is 0.003), one with a PTE~$> 0.99$,
and the rest appear
uniformly distributed between this range.  There are no apparent trends
or outliers among the various subsets or splits.

\begin{table*}
\caption{\label{tab:chi2}$\chi^2$ consistency tests for a selection of data splits and subsets}
\small%
\begin{center}
\begin{tabular}{llrrrrr}
\hline
\hline
 Temperature Data & & & \multicolumn{2}{c}{Difference} & \multicolumn{2}{c}{Sum} \\ 

 Split Type & Subset & \# DOF & $\chi^2$ & PTE & $\chi^2$ & PTE \\ 
\hline
Year                          	&	full                          	&		 1448	&	1474.2		&	0.31	&	        23188.7	&	$< 1 \times 10^{-16}$ \\ 
				&	s/n $>1$                      	&		  320	&	337.1		&	0.24	&	        21932.2	&	$< 1 \times 10^{-16}$ \\ 
				&	$l$ range 0--245              	&		  184	&	202.6		&	0.17	&	        10566.3	&	$< 1 \times 10^{-16}$ \\ 
				&	$l$ range 0--245 high s/n     	&		   36	&	 38.2		&	0.37	&	        10355.1	&	$< 1 \times 10^{-16}$ \\ 
				&	$l$ range 245--420            	&		  398	&	389.7		&	0.61	&	         7676.0	&	$< 1 \times 10^{-16}$ \\ 
				&	$l$ range 245--420 high s/n   	&		   79	&	 88.9		&	0.21	&	         7294.4	&	$< 1 \times 10^{-16}$ \\ 
				&	$l$ range 420--596            	&		  422	&	410.5		&	0.65	&	         3122.5	&	$< 1 \times 10^{-16}$ \\ 
				&	$l$ range 420--596 high s/n   	&		   84	&	 73.5		&	0.79	&	         2727.8	&	$< 1 \times 10^{-16}$ \\ 
				&	$l$ range 596--772            	&		  336	&	367.8		&	0.11	&	         1379.5	&	$< 1 \times 10^{-16}$ \\ 
				&	$l$ range 596--772 high s/n   	&		   67	&	 82.3		&	0.10	&	          991.8	&	$< 1 \times 10^{-16}$ \\ 
				&	$l$ range 772--1100           	&		  108	&	103.7		&	0.60	&	          444.4	&	$< 1 \times 10^{-16}$ \\ 
				&	$l$ range 772--1100 high s/n  	&		   21	&	 22.2		&	0.39	&	          307.7	&	$< 1 \times 10^{-16}$ \\ 
Epoch                         	&	full                          	&		 1520	&	1546.3		&	0.31	&	        32767.2	&	$< 1 \times 10^{-16}$ \\ 
				&	s/n $>1$                      	&		  348	&	366.5		&	0.24	&	        31430.0	&	$< 1 \times 10^{-16}$ \\ 
Azimuth range                 	&	full                          	&		 1520	&	1542.6		&	0.34	&	        32763.8	&	$< 1 \times 10^{-16}$ \\ 
				&	s/n $>1$                      	&		  348	&	355.2		&	0.38	&	        31426.9	&	$< 1 \times 10^{-16}$ \\ 
Faceplate position                 	&	full                          	&		 1318	&	1415.2		&	0.03	&	        27446.5	&	$< 1 \times 10^{-16}$ \\ 
				&	s/n $>1$                      	&		  331	&	365.3		&	0.09	&	        26270.1	&	$< 1 \times 10^{-16}$ \\ 
Stokes state            	&	full                          	&		 1524	&	1556.6		&	0.27	&	        33050.6	&	$< 1 \times 10^{-16}$ \\ 
				&	s/n $>1$                      	&		  350	&	358.2		&	0.37	&	        31722.5	&	$< 1 \times 10^{-16}$ \\ 
\hline
\hline
 Polarization Data & & & \multicolumn{2}{c}{Difference} & \multicolumn{2}{c}{Sum} \\ 

 Split Type & Subset & \# DOF & $\chi^2$ & PTE & $\chi^2$ & PTE \\ 
\hline
Year                          	&	full                          	&		 2896	&	2949.4		&	0.24	&	         2925.2	&	0.35 \\ 
				&	s/n $>1$                      	&		   30	&	 34.4		&	0.27	&	           82.4	&	$8.7\times 10^{-7}$ \\ 
				&	$l$ range 0--245              	&		  368	&	385.9		&	0.25	&	          315.0	&	0.98 \\ 
				&	$l$ range 0--245 high s/n     	&		   73	&	 61.0		&	0.84	&	           64.5	&	0.75 \\ 
				&	$l$ range 245--420            	&		  796	&	862.2		&	0.05	&	          829.4	&	0.20 \\ 
				&	$l$ range 245--420 high s/n   	&		  159	&	176.0		&	0.17	&	          223.8	&	$5.4\times 10^{-4}$ \\ 
				&	$l$ range 420--596            	&		  844	&	861.0		&	0.33	&	          837.3	&	0.56 \\ 
				&	$l$ range 420--596 high s/n   	&		  168	&	181.3		&	0.23	&	          189.7	&	0.12 \\ 
				&	$l$ range 596--772            	&		  672	&	648.1		&	0.74	&	          704.4	&	0.19 \\ 
				&	$l$ range 596--772 high s/n   	&		  134	&	139.5		&	0.35	&	          160.0	&	0.06 \\ 
				&	$l$ range 772--1100           	&		  216	&	192.3		&	0.88	&	          239.1	&	0.13 \\ 
				&	$l$ range 772--1100 high s/n  	&		   43	&	 32.3		&	0.88	&	           47.6	&	0.29 \\ 
Epoch                         	&	full                          	&		 3040	&	2907.1		&	0.96	&	         3112.2	&	0.18 \\ 
				&	s/n $>1$                      	&		   34	&	 29.2		&	0.70	&	           98.6	&	$3.3\times 10^{-8}$ \\ 
Azimuth range                 	&	full                          	&		 3040	&	3071.1		&	0.34	&	         3112.9	&	0.17 \\ 
				&	s/n $>1$                      	&		   34	&	 38.7		&	0.27	&	           98.7	&	$3.3\times 10^{-8}$ \\ 
Faceplate position                 	&	full                          	&		 2636	&	2710.4		&	0.15	&	         2722.2	&	0.12 \\ 
				&	s/n $>1$                      	&		   32	&	 43.6		&	0.08	&	           97.5	&	$1.6\times 10^{-8}$ \\ 
\hline

\end{tabular}
\end{center}
{\scshape Note} -- tabulated above are $\chi^2$ values for a representative selection  
of splits and subsets of the combined 2001/2002 dataset. Visibility
data containing the same sky signal is split to form two data vectors;
the $\chi^2$ statistic is then calculated on both the difference and
sum data vectors. Also tabulated are the number of degrees of freedom
(\# DOF), and probability to exceed (PTE) the value in the $\chi^2$
cumulative distribution function, to show the significance of the
result (PTE values indicated as $< 1 \times 10^{-16}$ are zero to the
precision with which we calculate the $\chi^2$ cumulative distribution
function). Difference data $\chi^2$ values test for systematic effects
in the data, while comparisons with sum data values test for the
presence of a repeatable sky signal. Temperature (co-polar) data are visibility
data in which the polarizers from both receivers are in the left ($LL$
Stokes state) or right ($RR$ Stokes state) circularly polarized state;
polarization (cross-polar) data are those in which the polarizers are in
opposite states ($LR$ or $RL$ Stokes state). In the temperature data, $LL$
and $RR$ Stokes state data are combined in all but the last type of
split. The s/n~$>1$ subset is the subset of s/n eigenmodes $>1$ and the
$l$ range high s/n subsets are the 20\% highest s/n modes. See
\S\ref{sec:chi2tests} for further description of the data split types
and subsets.  We have calculated 296 $\chi^2$ values for various split
types and subsets, with no obvious trends that would indicate
systematic contamination of the data.

\normalsize
\end{table*}

The high s/n mode subsets are more sensitive to certain classes of systematic effects in
the difference data vector and more sensitive to the expected sky signal in the sum data
vector, that otherwise may be masked by noise. Also, the number of
modes with s/n~$> 1$ gives an indication of the expected power of the
experiment to constrain the sky signal. The co-polar data, which are
sensitive to the temperature signal, have many more high s/n modes
than the cross-polar data, which are only sensitive to polarized
radiation.  Within the context of the concordance model used to
generate the s/n eigenmode basis, we have sensitivity with an expected s/n~$>1$ to
$\sim 340$ temperature (co-polar) modes vs. $\sim 34$ polarization
(cross-polar) modes.

\subsection{Detection of Signal}

\begin{figure*}[t]
\begin{center}
\epsfig{file=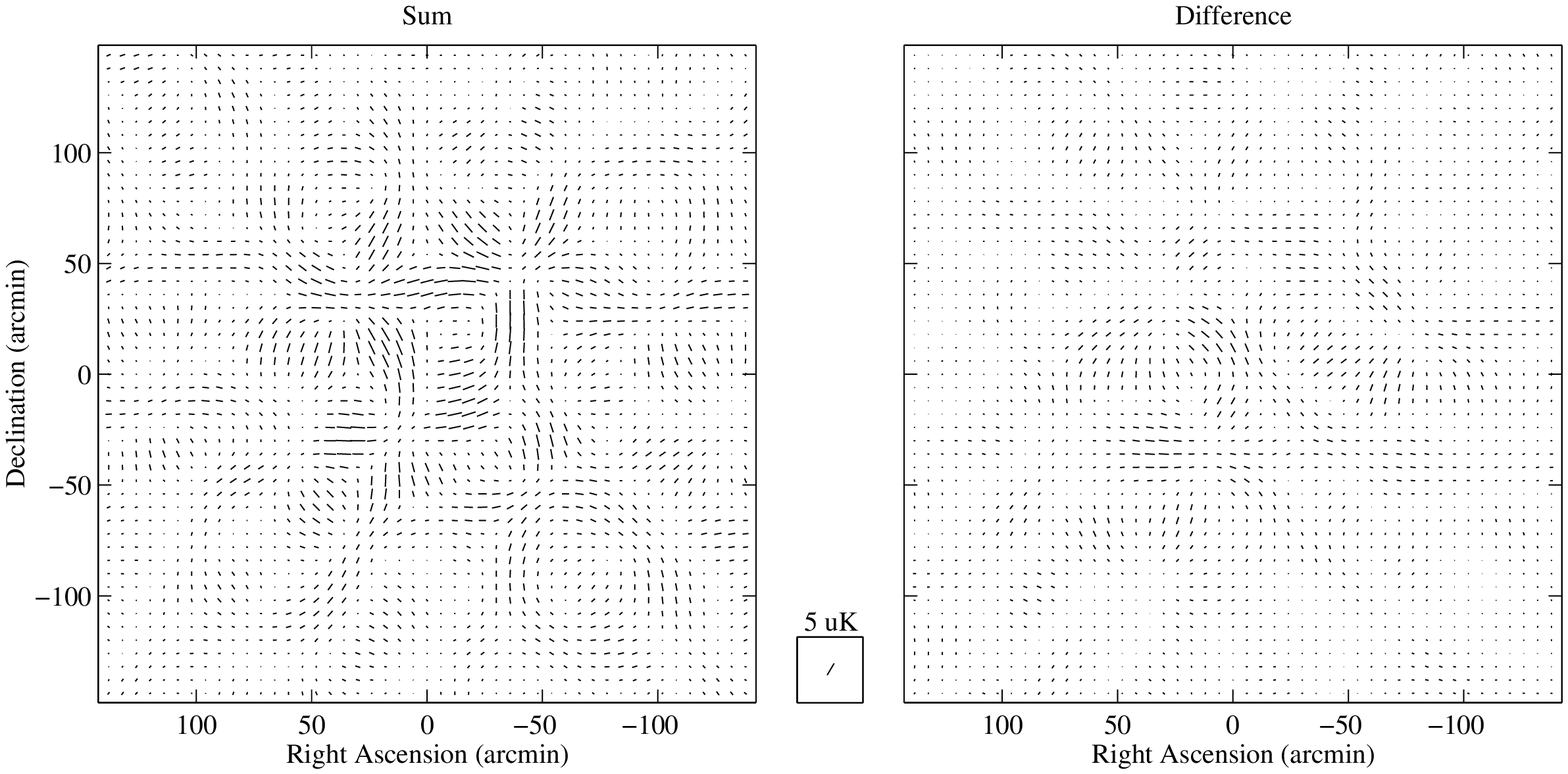,width=6.5in}
\end{center}
\caption{Polarization maps constructed from polarized datasets that have
been split by epoch, and formed into sum (left) or difference (right) 
data vectors, as reported in \S\ref{sec:chi2tests}.
In order to isolate the most significant
signal in our data, we have used only the subset of 34 eigenmodes which, under the
concordance model, are expected to have average signal/noise $>1$.
Unlike conventional interferometer maps, the signal/noise 
selected eigenmodes reflect the gain of the primary beam.  This is
apparent in the difference map (right), which is consistent with noise.
Comparison of this map to the sum map (left) illustrates a result also given
numerically for this split/subset in Table~\ref{tab:chi2}: 
that these individual modes in the polarized dataset show a
significant signal.}
\label{fig:polmap}
\end{figure*}

Given that the data show remarkable consistency in $\chi^2$ tests of
the difference data vectors, the $\chi^2$ values of the sum data
vectors can be used to test for the presence of sky signal, independently
of likelihood analysis methods described below in \S\ref{sec:lhformalism}. In the
co-polar data, all splits and subsets show highly significant $\chi^2$
values (PTE~$< 1\times 10^{-16}$, the precision to which we calculate
the cumulative distribution function).

For the cross-polar data, the sum
data vector $\chi^2$ values for the high s/n subsets show high
significance, with the PTE $< 1 \times 10^{-6}$ for all s/n~$>1$
subsets in Table~\ref{tab:chi2}.  This simple and powerful test
indicates that we have detected, with high significance, the presence
of a polarized signal in the data, and that this signal is
repeatable in all of the data splits.
It should be noted that at this stage in the analysis, the data have not 
been corrected for the off axis leakage described in Paper IV. 
In the following likelihood analysis, the off axis leakage is included and shown
to have an insignificant impact on the results. 
The polarization map shown in Figure~\ref{fig:polmap} gives a visual
representation of this repeatable polarization signal.
Shown are the epoch split sum and difference polarization maps constructed using only the 34
highest s/n modes, as formed in the concordance model s/n eigenmode basis.
The sum map shows a repeatable polarized signal, while the
difference map is consistent with instrument noise.

The likelihood analysis described in the following sections makes use of all of the
information in our dataset.
Such an analysis, in principle, may yield statistically significant evidence of a signal
even in cases of datasets for which it is not possible to isolate any individual modes
which have average s/n $>1$.  However, the existence of such modes in our dataset, which has resulted
from our strategy of integrating deeply on a limited patch of sky, allows us
to determine the presence of the signal with the very simple analysis described above.
It also reduces sensitivity to the noise model estimation in the  likelihood results that we report next.  Finally, it gives our dataset greater power to exclude the possibility
of no signal than it might have had if we had observed more modes but with less s/n in each.

\section{Likelihood Analysis Formalism}
\label{sec:lhformalism}

The preceding section gives strong evidence for the presence of a signal in our polarization
data. We now quantify the amplitude of that signal using the
standard tool 
of likelihood
analysis. In such an analysis, the covariance of the signal, 
$\mathbf{C}_{T}\left(\mathbf{\kappa}\right)$, is modeled in
terms of parameters $\mathbf{\kappa}$ appropriate for describing the temperature and
polarization anisotropies of the CMB. The covariance of the data vector is
modeled $\mathbf{C}\left(  \mathbf{\kappa}\right)  \equiv\mathbf{C}_{T}\left(
\mathbf{\kappa}\right)  +\mathbf{C}_{N}$, where $\mathbf{C}_{N}$ is the noise
covariance matrix. Given our data vector $\mathbf{\Delta}$, the likelihood
of the model specified by the parameter vector $\mathbf{\kappa}$ is
the probability of our data vector given that model,
\begin{eqnarray}
\label{eq:like}
\nonumber
L\left(  \mathbf{\kappa}\right) &=& P\left( \mathbf{\Delta} | \mathbf{\kappa} \right)  \\
&\propto& \det\left(  \mathbf{C}\left(
\mathbf{\kappa}\right)  \right)  ^{-1/2}\exp\left(  -\frac{1}{2}%
\mathbf{\Delta}^{t}\mathbf{C}\left(  \mathbf{\kappa}\right)  ^{-1}%
\mathbf{\Delta}\right)
\end{eqnarray}

Although the full likelihood function itself is the most basic result of the
likelihood analysis, it is useful to identify and report  
the values of the parameters that maximize the
likelihood (so-called \emph{maximum likelihood (ML) estimators}).
Uncertainties in the parameter values can be
estimated by characterizing the shape of the likelihood surface,
as discussed further in \S\ref{sec:lreport}.

\subsection{The CMB Power Spectra}

The temperature and polarization anisotropies of the CMB can be 
characterized statistically by six angular power spectra: three that give the
amplitudes of temperature, $E$-mode and $B$-mode polarization anisotropies
as a function of angular scale, and three that describe correlations
between them.  These spectra are written $C_{l}^{X}$, with $X =
\left\{T,E,B,TE,TB,EB \right\}$.  In our likelihood analyses, we
choose various parameterizations of these spectra to constrain.

For a given cosmological model, these spectra can be readily
calculated using efficient, publicly-available Boltzmann codes
\markcite{zaldarriaga00}({Zaldarriaga} \& {Seljak} 2000).
Details of how to define these spectra in terms of all-sky multipole expansions of the
temperature and linear polarization of the CMB radiation field 
are given by \markcite{zaldarriaga97}{Zaldarriaga} \& {Seljak} (1997) and \markcite{kamionkowski97}Kamionkowski {et~al.} (1997).
For DASI's $3.4^\circ$ field of view, a flat sky approximation is
appropriate \markcite{white99a}(White {et~al.} 1999), so that the spectra may be defined
somewhat more simply.  In this approximation
the temperature angular power spectrum
is defined
\begin{equation}
C_{l}^{T}\simeq C^{T}\left( \left\vert \mathbf{u}\right\vert \right)  \equiv\left\langle \frac{{\widetilde
{T}}^\ast\! \left(  \mathbf{u} \right) \widetilde{T}\left(
\mathbf{u}\right)  }{T^{2}_{\mathrm{CMB}}}\right\rangle,
\end{equation}
where $\widetilde{T}(\mathbf{u})$ is the Fourier transform of
$T(\mathbf{x})$, $T_{\mathrm{CMB}}$ is the mean temperature of the
CMB, and $l/2\pi =  \left\vert \mathbf{u}\right\vert $ gives
the correspondence between multipole $l$ and Fourier radius $\left\vert \mathbf{u}\right\vert$.
The other spectra in the flat sky approximation are similarly
defined, e.g., $C^{TE}\left( u\right) \equiv\left\langle
{\widetilde{T}}^\ast\!\left( \mathbf{u} \right) \widetilde{E}\left(
\mathbf{u}\right) /T^{2}_{\mathrm{CMB}}\right\rangle $.  
The relationship between $\widetilde{E},\widetilde{B}$ and the linear polarization
Stokes parameters $Q$ and $U$ is
\begin{eqnarray}
\nonumber
\widetilde{Q}\left(  \mathbf{u}\right)    &=& \cos\left(  2\chi\right)
\widetilde{E}\left(  \mathbf{u}\right)  -\sin\left(  2\chi\right)
\widetilde{B}\left(  \mathbf{u}\right)  \nonumber \\ 
\widetilde{U}\left(  \mathbf{u}\right)    &=& \sin\left(  2\chi\right)
\widetilde{E}\left(  \mathbf{u}\right)  +\cos\left(  2\chi\right)
\widetilde{B}\left(  \mathbf{u}\right).
\label{eq:queb}
\end{eqnarray}
where $\chi = \arg(\mathbf{u})$ and the polarization orientation angle
defining $Q,U$ are both measured on the sky from north through east.

\subsection{Theory Covariance Matrix}
\label{sec:theory}

The theory covariance matrix is the expected covariance of
the signal component of the datavector,
$\mathbf{C}_{T} \equiv \left<\mathbf{s}\mathbf{s}^t\right>$.
The signals measured by the visibilities in our datavector for a given
baseline $\mathbf{u}_{i}$ (after calibration and leakage correction) are
\begin{eqnarray}
\label{eq:vis}
\nonumber
V^{RR}\left(  \mathbf{u}_{i}\right)   &=& \alpha_{i}\int d\mathbf{x\ }A\left(
\mathbf{x},\nu_{i}\right)  \left[  T\left(  \mathbf{x}\right)
+V\left(  \mathbf{x}\right)  \right]  e^{-2\pi i\mathbf{u}_{i}\mathbf{\cdot
x}}\\\nonumber
V^{LL}\left(  \mathbf{u}_{i}\right)   &=& \alpha_{i}\int d\mathbf{x\ }A\left(
\mathbf{x},\nu_{i}\right)  \left[  T\left(  \mathbf{x}\right)
-V\left(  \mathbf{x}\right)  \right]  e^{-2\pi i\mathbf{u}_{i}\mathbf{\cdot
x}}\\\nonumber
V^{RL}\left(  \mathbf{u}_{i}\right)   &=& \alpha_{i}\int d\mathbf{x\ }A\left(
\mathbf{x},\nu_{i}\right)  \left[  Q\left(  \mathbf{x}\right)  +iU\left(
\mathbf{x}\right)  \right]  e^{-2\pi i\mathbf{u}_{i}\mathbf{\cdot x}}\\\nonumber
V^{LR}\left(  \mathbf{u}_{i}\right)   &=& \alpha_{i}\int d\mathbf{x\ }A\left(
\mathbf{x},\nu_{i}\right)  \left[  Q\left(  \mathbf{x}\right)  -iU\left(
\mathbf{x}\right)  \right]  e^{-2\pi i\mathbf{u}_{i}\mathbf{\cdot x}},\\
\end{eqnarray}
where $A\left(  \mathbf{x},\nu_{i}\right)  $ specifies the beam power pattern
at frequency $\nu_{i}$, $T(\mathbf{x})$, $Q(\mathbf{x})$, $U(\mathbf{x})$, and
$V(\mathbf{x})$ are the four Stokes parameters in units of CMB temperature ($\mu K$),
and $\alpha_{i}=\partial B_{\text{Planck}}\left(\nu_{i},T_{\mathrm{CMB}}\right)  /\partial T$
is the appropriate factor for converting from
these units to flux density ($Jy$).  The co-polar
visibilities $V^{RR}$ and $V^{LL}$ are sensitive to the Fourier
transform of the temperature signal $T({\bf x})$ and circular polarization
component $V({\bf x})$ (expected to be zero).  The cross-polar
visibilities $V^{RL}$ and $V^{LR}$ are sensitive to the Fourier
transform of the linear polarization components $Q,U$.
Using Equation~\ref{eq:queb}, it can be seen that pairwise combinations of the
visibilities are direct measures of nearly pure $T$, $E$ and $B$ Fourier modes
on the sky, so that the dataset easily lends itself to placing independent constraints
on these power spectra.

We construct the theory covariance matrix as the sum of components for each
parameter in the analysis
\begin{equation}
\mathbf{C}_{T}\left(\kappa\right)  = \sum_{p}\kappa_{p}\mathbf{B}_{T}^{p}.
\end{equation}
From Equations 2 -- 4, it is possible to derive a general
expression for the matrix elements of a theory matrix component,
\begin{eqnarray}
\nonumber
B_{T\ ij}^{p}&=&\frac{1}{2}\alpha_{i}\alpha_{j}T^{2}_{\mathrm{CMB}}\int d\mathbf{u}
\ C^{X}\left(  u\right)  \widetilde{A}\left(  \mathbf{u}_{i}-\mathbf{u}
,\nu_{i}\right)\\
&\times& \left[  \zeta_{1}\widetilde{A}\left(  \mathbf{u}
_{j}-\mathbf{u},\nu_{j}\right)  +\zeta_{2}\widetilde{A}\left(  \mathbf{u}
_{j}+\mathbf{u},\nu_{j}\right)  \right].
\end{eqnarray}
The coefficients $\zeta_{1}$ and $\zeta_{2}$ can take values $\left\{
0,\pm1,\pm2\right\}  \times\left\{  \cos\left\{  2\chi,4\chi\right\}
,\sin\left\{  2\chi,4\chi\right\}  \right\}  $ depending on the Stokes states
$(RR,LL,RL,LR)$ of each of the two baselines $i$ and $j$ and on which
of the six spectra $(T,E,B,TE,TB,EB)$\ is specified by $X$.
The integration may be limited to annular regions which correspond to
$l-$ranges over which the power spectrum $C^X$ is hypothesized to be relatively
flat, or else some shape of the spectrum may be postulated.

Potentially contaminated modes in the data vector may be effectively
projected out using a constraint matrix formalism \markcite{bond97}({Bond} {et~al.} 1998).
This formalism can be used to remove the effect of point sources of
known position without knowledge of their flux densities, as 
described in Paper II. This procedure can be generalized
to include the case of polarized point sources. Although we
have tested for the presence of point sources in the polarization
power spectra using this method, in the final analysis we use constraint
matrices to project point sources out of the temperature data only, and
not the polarization data (see \S\ref{sec:pointsources}).

The off-axis leakage, discussed in \S\ref{sec:offaxisleak} and in
detail in Paper IV, has the effect of mixing some power from the
temperature signal $T$ into the cross-polar visibilities.
Our model of the off-axis leakage allows us write an expression for
it analogous to Equation \ref{eq:vis}, and to construct a corresponding
theory covariance matrix component to account for it.
In practice, this is a small effect, as discussed in \S\ref{sec:systematics}.

\subsection{Likelihood Parameters}
\label{sec:lpar}

In \S\ref{sec:lhresults} we present the results from
nine separate likelihood analyses, organized in three groups:
analyses using the polarization (cross-polar) data only, using the
temperature (co-polar) data only, and using the full joint temperature
and polarization dataset.
Our choice of parameters with which to characterize the
six CMB power spectra is a compromise between maximizing 
sensitivity to the signal and constraining the shape of the power
spectra.  
In the different analyses we either characterize various power spectra with
a single amplitude parameter covering all angular scales, or split the $l-$range into five bands
over which spectra are approximated as piecewise-flat, in units of
$l(l+1)C_{l}/(2\pi)$.  Five bands were chosen as a compromise
between too many for the data to bear and too few to capture the shape
of the underlying power spectra.  The $l-$ranges of these five bands
are based on those of the nine-band analysis of Paper II; we have
simply combined the first four pairs of these bands, and kept the ninth
as before.
In some analyses we also constrain the frequency spectral
indices of the temperature and polarization power spectra as a test
for foreground contamination.  

The $l-$range to which DASI has non-zero sensitivity is $28 < l < 1047$.
That range includes the first three peaks of the temperature power spectrum,
and within it the amplitude of that spectrum,
which we express in units $l(l+1)C_{l}/(2\pi)$, varies by
a factor of $\sim 4$.  Over this same range, the $E$-mode polarization spectrum is predicted
to have four peaks while rising roughly as $l^2$ (in the same units),
varying in amplitude by nearly two orders of magnitude \markcite{hu_w97}(Hu \& White 1997).
The $TE$ correlation is predicted to exhibit a complex spectrum that in fact crosses 
zero five times in this range.

For the single bandpower analyses which maximize our sensitivity
to a potential signal, the shape of the model power spectrum assumed
will have an effect on the sensitivity of the result.  In particular,
if the assumed shape is a poor fit to the true spectrum preferred by the data,
the results will be both less powerful and difficult to interpret.
For temperature spectrum measurements, the most common choice in recent years
has been the so-called flat bandpower, $C_l \propto 1/l(l+1)$, which matches
the gross large-scale power law shape of that spectrum.
Because of extreme variations predicted in the $E$ and $TE$ spectra over DASI's $l-$range,
we do not expect a single flat bandpower
parameterization to be a good description of the data.
In fact, a more appropriate definition
of ``flat bandpower'' for polarization measurements sensitive to large ranges of
$l < 1000$ might be $C_l \propto \rm{const}$.  Other shapes have been tried,
notably the Gaussian autocorrelation function (by the PIQUE group \markcite{hedman01}({Hedman} {et~al.} 2001))
which reduces to $C_l \propto \rm{const}$ at large scales and perhaps offers
a better fit to the gross amplitude of the predicted $E$ spectrum.  

In our single band analyses, we have chosen a shape for our single bandpower
parameters based on the predicted spectra for a cosmological model currently
favored by observations.  The specific model that we choose---which we will call
the {\it concordance model}---is a $\Lambda$CDM model with flat spatial curvature, 5\% baryonic
matter, 35\% dark matter, 60\% dark energy, and a Hubble constant of
65~km~s$^{-1}$~Mpc$^{-1}$, $(\Omega_b = 0.05, \Omega_{cdm} = 0.35,
\Omega_\Lambda = 0.60, h = 0.65)$ and the exact normalization
$C_{10} = 700 \mu K^2$.  This concordance model was defined in Paper III as a
good fit to the DASI temperature power spectrum and other observations.
The concordance model spectra for $T$, $E$, and $TE$ are shown in Figure~\ref{fig:fivebandTEBTE}.
The five flat bandpower likelihood results shown in the same figure, and discussed
in the next section, suggest that the concordance shaped spectra do indeed better 
characterize
the data than any power-law approximation.  In \S\ref{sec:eb_anal}, we explicitly
test the likelihood of the concordance model parameterization against that of the
two power laws mentioned above, and find that the concordance model shape is strongly
preferred by the data.

\begin{figure*}[th]
\begin{center}
\epsfig{file=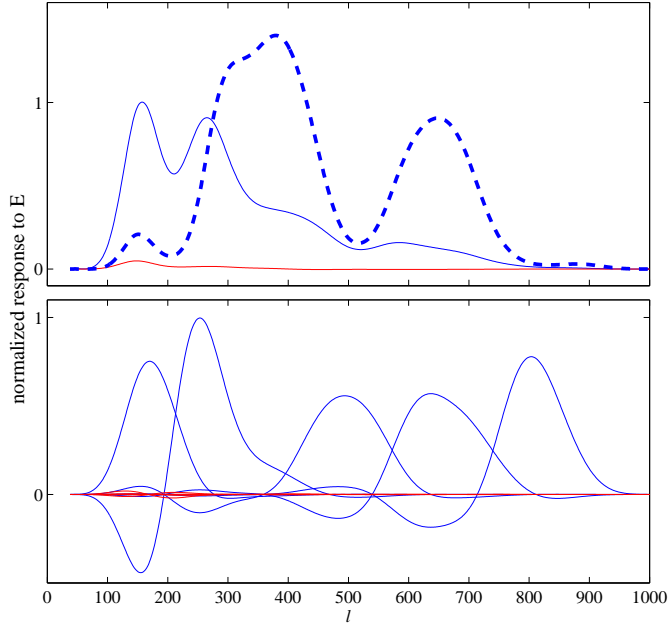,width=3.5in}
\end{center}
\caption{The upper panel shows the $E$ (solid blue) and B (solid red, much lower curve)
parameter window functions that indicate the response to the $E$ power spectrum of the
two parameters in our $E$/B analysis.
The blue dashed curve shows the result of multiplying the $E$ parameter window
function by the concordance $E$ spectrum, illustrating that for this CMB spectrum,
most of the response of our experiment's $E$ parameter comes from the region of the second
peak ($250 \lesssim l \lesssim 450$), with a substantial contribution also from the
third peak and a smaller contribution from the first.
The lower panel shows $E1$ -- $E5$ (blue) and $B1$ -- $B5$ (red, again much lower) parameter window functions for
the $E$ power spectrum from our $E5$/$B5$ analysis.  DASI's response to $E$ and B is very symmetric, so that the
corresponding plots that show these parameters response to the B power spectrum
are nearly identical to these, with the $E$ and B parameters reversed.}
\label{fig:windowfunc}
\end{figure*}

It should be noted that the likelihood analysis is always model
dependent, regardless of whether a flat or shaped model is chosen for
parameterization.  To evaluate the expectation value of the results for a
hypothesized theoretical power spectrum, one must use window functions
appropriate for the parameters of the particular analysis.
The calculation of such parameter window functions has been described by 
\markcite{knox99}{Knox} (1999), \markcite{Halverson_thesis}{Halverson} (2002), and in particular for polarization
spectra by \markcite{tegmark01}{Tegmark} \& {de Oliveira-Costa} (2001).
In general, the parameter window function has a non-trivial shape (even for
a flat band-power analysis) which is dependent on the shape of
the true spectra as well as the intrinsic sensitivity of the
instrument as a function of angular scale. The parameter window
functions for the $E/B$ polarization analysis are shown in
Figure~\ref{fig:windowfunc}, and are also available on our
website\footnote{\texttt http://astro.uchicago.edu/dasi}.  

\subsection{Likelihood Evaluation}

Prior to likelihood analysis, the data vector and the covariance
matrices can be compressed by combining visibility data from nearby
points in the $(u,v)$ plane, where the signal is highly
correlated.  This reduces the computational time required for the
analyses without a significant loss of information about the signal.
All analyses were run on standard desktop workstations.

For each analysis, we use an iterated quadratic estimator technique to
find the ML values of our parameters \markcite{bond97}({Bond} {et~al.} 1998).  We also
explicitly map out the likelihood function by evaluating Equation
\ref{eq:like} over a uniform parameter grid large enough to enclose
all regions of substantial likelihood.  A single
likelihood evaluation typically takes several seconds, so this
explicit grid evaluation is impractical for the analyses
which include five or more parameters.
For each analysis we also implement a Markov chain evaluation of the likelihood
function \markcite{christensen01}({Christensen} {et~al.} 2001).  We find this to be a useful and efficient tool
for mapping the likelihoods of these high-dimensional parameter spaces in the
region of substantial likelihood.  We have compared the Markov technique
to the grid evaluation for the lower-dimensional analyses and found
the results to be in excellent agreement.  In all cases, the peak of the
full likelihood evaluated with either technique is confirmed to
coincide with the ML values returned by the iterated quadratic estimator.

\subsection{Simulations and Parameter Recovery Tests}

The likelihood analysis software was
extensively tested through analysis of simulated data.  The analysis
software and data simulation software were independently authored,
as a check for potential coding errors.

Simulated sky maps were generated from realizations of a variety of
smooth CMB power spectra, including both the concordance spectrum and
various non-concordance models, both with and without $E$ and $B$ polarization and
$TE$ correlations.  Independent realizations of the sky were
``observed'' to construct simulated visibilities with Fourier-plane
sampling identical to the real data.  The simulations were
designed to replicate the actual data as realistically as possible
and include artifacts of the instrumental polarization response and
calibration, such as the on-axis and off-axis leakages described in
\S\ref{sec:instrpol}, and the cross-polar phase offset described in
\S\ref{sec:gainphase}, allowing us to test the calibration and treatment
of these effects implemented in the analysis software.

Each of the analyses described in \S\ref{sec:lhresults} was performed
on hundreds of these simulated datasets with independent realizations
of sky and instrument noise, both with noise variances that matched
the real data, and with noise a factor of 10 lower.  In all cases, we
found that the means of the ML estimators recovered the expectation
values $\left\langle \kappa_p \right\rangle$ of each parameter without
evidence of bias, and that the variance of the ML
estimators was found to be consistent with the estimated uncertainty given by $F^{-1}$
evaluated at $\left\langle \mathbf{\kappa}\right\rangle$, where $F$ is
the Fisher matrix.

\subsection{Reporting of Likelihood Results}
\label{sec:lreport}

Likelihood results reported in this paper are the global maxima of the
multidimensional likelihood functions.  Confidence intervals are
determined by integrating the likelihood over the full parameter grid,
including non-physical values; the reported intervals are the
equal-likelihood bounds which enclose $68\%$ of the total probability.
This prescription corresponds to what is generally referred to as the
{\it highest posterior density} (HPD) interval.  Results for single
parameters are obtained by integrating (marginalizing) the likelihood function over
the other parameters.  In the tabulated results, we also report
marginalized uncertainties obtained by evaluating the Fisher matrix
at the maximum likelihood model, i.e., $\left(F^{-1}\right)
_{ii}^{1/2}$.  Although in most cases, the two confidence intervals
are quite similar, we regard the HPD interval as the primary result.

For parameters which are intrinsically positive 
we also marginalize the likelihood distribution after imposing a prior
that excludes the unphysical negative values.
We then test if the 95\% integral point has a likelihood smaller
than that at zero; if it does the confidence interval should be
regarded as an upper limit rather than a detection and
we quote the corresponding value.

\subsection{Goodness-of-Fit Tests}
\label{sec:goodfit}

Using the likelihood function, we wish to determine if our results are consistent
with a given model.  For example, we would like to know the significance of 
any detections and determine if the polarization data are consistent 
with that predicted in a given cosmological model.
We define as a goodness-of-fit statistic the logarithmic ratio of the 
maximum of the likelihood to its value for some model ${\cal H}_0$ described by parameters
$\kappa_{0}$.
\[
\Lambda({\cal H}_0)\equiv-\log\left(  \frac{L\left(  \kappa_{ML}\right)  }{L\left(
\kappa_{0}\right)  }\right)\,.
\]
The statistic $\Lambda$ simply indicates how much the likelihood has fallen
from its peak value down to its value at $\kappa_{0}$.  Large values indicate
inconsistency of the likelihood result with the model ${\cal H}_0$.
To assess significance, we perform Monte Carlo (MC) simulations of this
statistic under the hypothesis that ${\cal H}_{0}$ is true.
From this, we can determine the probability, given ${\cal H}_0$ true, to obtain a value
of $\Lambda$ that exceeds the observed value, which we hereafter refer to as PTE. 

When considering models which the data indicate to be very unlikely, sufficient 
sampling of the likelihood statistic becomes computationally prohibitive; 
our typical MC simulations are limited to only 1000 realizations. 
In the limit that the parameter errors are normally distributed, 
our chosen statistic reduces to $\Lambda=\Delta\chi^2/2$.
The integral over the $\chi^2$ distribution is described by an incomplete 
gamma function;
\[
{\rm PTE}=\frac{1}{\Gamma(N/2)} {\int^{\infty}_{\Lambda} e^{-x} x^{\frac{N}{2}-1} dx}
\]
where $\Gamma(x)$ is the complete gamma function, and N is the number of parameters.
Neither the likelihood function nor the distribution of the ML
estimators is, in general, normally distributed, and therefore this approximation
must be tested.  In all cases where we can compute a meaningful PTE with MC simulations,
we have done so and found the results to be in excellent agreement with the
analytic approximation.  
Therefore, we are confident that adopting this approximation is justified. 
All results for PTE in this paper are calculated using this analytic
expression unless otherwise stated.

\section{Likelihood Results}

\label{sec:lhresults}

In the following sections we will first discuss analyses
based on polarization data only,
then those based on temperature data only, and finally joint
temperature-polarization analyses.
Numerical results for the analyses described in this section are given
in Tables \ref{tab:cxpolar_like}, \ref{tab:copolar_like} and
\ref{tab:joint_like}.
The parameter correlation matrices are tabulated in
Appendix~\ref{sec:corrtab}.  The conventions used for reporting likelihood
results have been discussed in \S\ref{sec:lreport}.

\subsection{Polarization Data Analyses and Results for $E$ and $B$ Parameters}

\subsubsection{$E/B$ Analysis}
\label{sec:eb_anal}

The E/B analysis uses two single bandpower parameters to characterize the amplitudes
of the $E$ and B polarization spectra.  As discussed in \S\ref{sec:lpar}, this analysis
requires a choice of shape for the spectra to be parameterized.  DASI has instrumental
sensitivity to $E$ and B that is symmetrical and nearly independent.  Although the B spectrum
is not expected to have the same shape
as the $E$ spectrum,
we choose the same shape for both spectra in order to make the analysis also symmetrical.

We have considered three {\it a priori} shapes to check which is most appropriate for our data:
the concordance $E$ spectrum shape (as defined in \S\ref{sec:lpar}), and
two ``power law'' alternatives, $C_l \propto 1/l(l+1)$ (commonly called ``flat'') and $C_l \propto \rm{const.}$.
For each of these three cases, the point at $E=0,B=0$ corresponds to the same
zero-polarization ``nopol'' model, so that the likelihood ratios $\Lambda(\rm{nopol})$
may be compared directly to assess the relative likelihoods of the best-fit model in
each case.  For the $C_l \propto 1/l(l+1)$ case, the ML values are $E=6.8 \mu K^2,
B=-0.4 \mu K^2$, with the log-likelihood at zero falling by $\Lambda(\rm{nopol}) = 4.34$.
For the $C_l \propto \rm{const.}$ case, the ML values are $E=5.1 \mu K^2,
B=1.2 \mu K^2$ at $l=300$, with $\Lambda(\rm{nopol}) = 8.48$.  For the concordance
shape, the ML values are $E=0.80,B=0.21$ in units of the concordance $E$ spectrum amplitude,
with $\Lambda(\rm{nopol}) = 13.76$.  The likelihood of the best fit model in the
concordance case is a factor of 200 and 12,000 higher than those of the $C_l \propto \rm{const.}$
and $C_l \propto 1/l(l+1)$ cases, respectively, and so compared to the concordance shape
either of these is a very poor model for the data.  The data clearly prefer the concordance
shape, which we use for our E/B and other single bandpower analyses.

\begin{figure*}[t]
\begin{center}
\epsfig{file=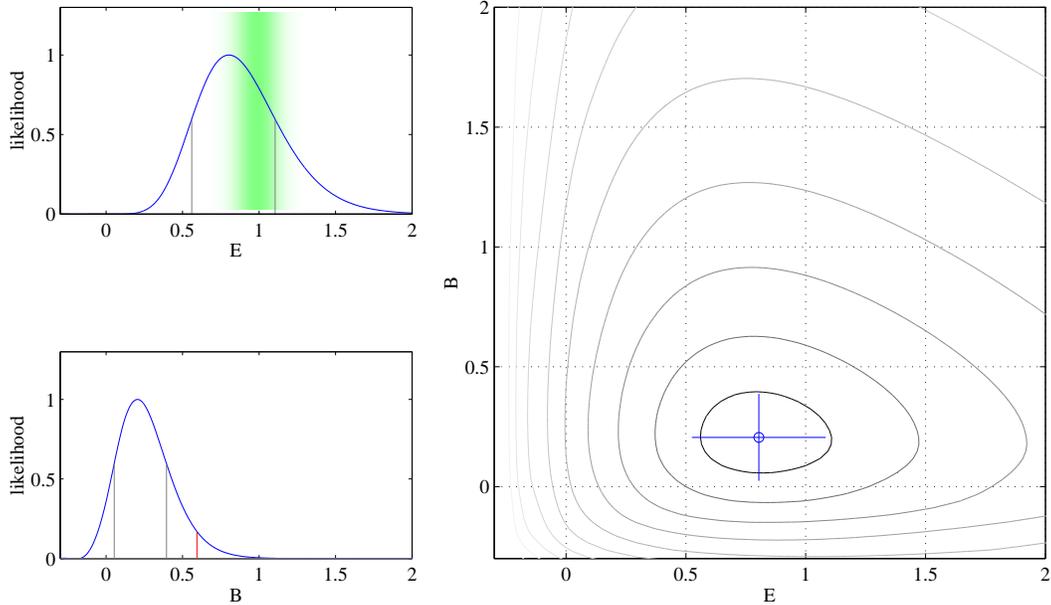,width=5.5in}
\end{center}
\caption{Results from the two parameter shaped bandpower E/B polarization
analysis assuming an $E$-mode power spectrum shape as predicted for
the concordance model, in units of amplitude relative to that model.
The same shape is assumed for the $B$-mode spectrum.
(right panel) The point shows the maximum likelihood value with
the cross indicating Fisher matrix errors.  Likelihood contours are placed at levels
$\exp(-n^2/2)$ relative to the maximum, i.e., for a normal distribution,
the extrema of these contours along either dimension would give the marginalized
$n$-sigma interval.
(left panels) The corresponding single parameter likelihood
distributions marginalized over the other parameter.
The grey lines enclose 68\% of the total likelihood.
The red line indicates the 95\% confidence upper limit on $B$-mode power.
The green band shows the distribution of $E$ expectation values
for a large grid of cosmological models weighted by the likelihood
of those models given our previous temperature result
(see Paper III).
}
\label{fig:EBplot}
\end{figure*}

Figure~\ref{fig:EBplot} illustrates the result of this E/B polarization analysis.
As stated above, we find that the maximum likelihood value of $E$ is 0.80 with a
68\% confidence interval of (0.56 to 1.10).
For B, the result should clearly be regarded as a upper limit;
95\% of the ${\rm B}>0$ likelihood (marginalized over $E$) lies below 0.59.

The upper panel of Figure~\ref{fig:windowfunc} shows the parameter
window functions relevant for this analysis.
Note that the $E$ parameter has very little sensitivity to B and
vice versa ---the purity with which DASI can separate these
is remarkable.  This is also demonstrated by the low correlation
($-0.046$) between the $E$ and B parameters as determined from the Fisher matrix,
as reported in Appendix \ref{sec:corrtab}.

Assuming that the uncertainties in $E$ and $B$ are normally distributed
(\S\ref{sec:lreport}), we 
estimate the probability that our data are consistent with the
zero polarization hypothesis to be $\rm{PTE} = 1.05 \times 10^{-6}$.
Our data are highly incompatible with the no polarization hypothesis.
Marginalizing over B, we find $\Lambda(E=0) = 12.1$ corresponding 
to detection of $E$-mode polarization at a PTE of $8.46 \times 10^{-7}$ (or 
a significance of $4.92 \sigma$).

The likelihood ratio for the concordance model 
gives $\Lambda(E=1,\,B=0)=1.23$, for which the Monte Carlo and analytic PTE are both 0.28.
We conclude that our data are consistent with the concordance model.

However, the temperature power spectrum of the CMB is still somewhat uncertain
and even within the $\sim 7$ parameter class of cosmological models often
considered, the shape and amplitude of the predicted $E$-mode spectrum is
still somewhat uncertain.
To quantify this, we have taken the model grid generated for Paper III
and calculated the expectation value of the shaped band $E$ parameter
for each model using the window function shown in
Figure~\ref{fig:windowfunc}.
We then take the distribution of these predicted $E$ amplitudes, weighted
by the likelihood of the corresponding model given our previous
temperature results (using a common calibration uncertainty
for the DASI temperature and polarization measurements).
This yields a 68\% credible interval for the predicted value of the
$E$ parameter of 0.90 to 1.11.
Our data are compatible with the expectation for $E$ based on existing knowledge 
of the temperature spectrum.

\label{sec:EB}

\subsubsection{E/$\beta_E$}

We have performed a two parameter analysis to determine
the amplitude of the $E$-mode polarization signal as above and
the frequency spectral index $\beta_E$ of this signal relative to
CMB (Figure~\ref{fig:specind}).
As expected, the results for the $E$-mode amplitude are very similar
to those for the E/B analysis described in the previous section.
The spectral index constraint is not strong; the maximum
likelihood value is $\beta_E = 0.17$ ($-1.63$ -- 1.92).
This result is nevertheless interesting in the context
of ruling out possible foregrounds (see \S\ref{sec:diffusefore} below).

\subsubsection{$E5$/$B5$}
\label{sec:E5B5}

\begin{figure*}[t]
\begin{center}
\epsfig{file=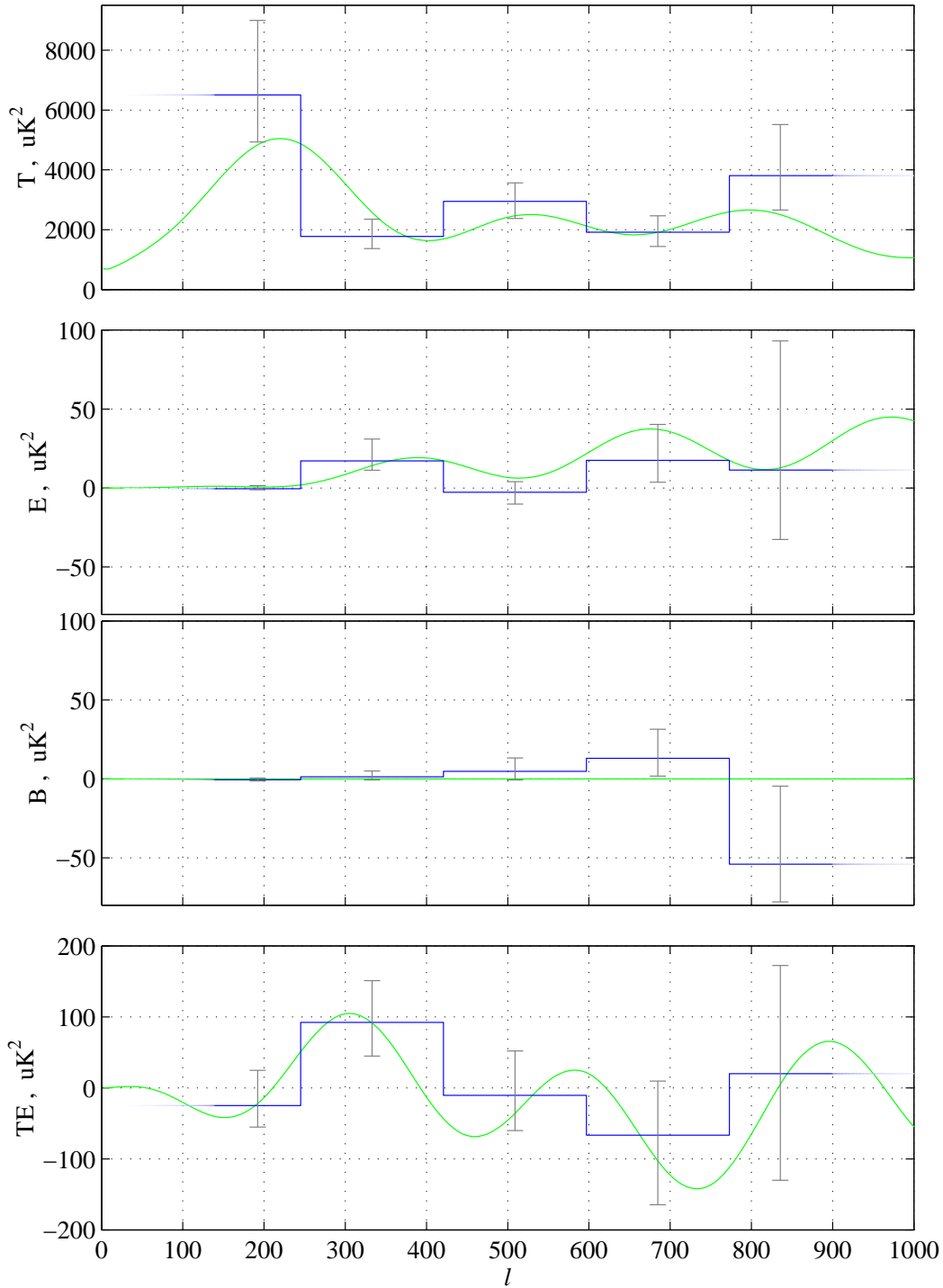,width=5.5in}
\end{center}
\caption{Results from several likelihood analyses: 
The T5 temperature analysis is shown in the top panel.
The ten parameter $E5$/$B5$ polarization analysis is shown in the middle two panels.
The 5 $TE$ bands from the $T$/$E$/$TE5$ joint analysis are shown in the bottom panel.
All the results shown are flat bandpower values.
The blue line shows the maximum likelihood bandpower values with the
grey error bars indicating the 68\% central region of the likelihood
marginalizing over the other parameter values (analogous to the grey lines
in Figure~\ref{fig:EBplot}).
In each case the green line is the concordance model.
}
\label{fig:fivebandTEBTE}
\end{figure*}

The central two panels of Figure~\ref{fig:fivebandTEBTE} show the
results of a ten parameter analysis characterizing the $E$ and $B$-mode spectra
using five flat bandpowers for each.
The lower panel of Figure~\ref{fig:windowfunc} shows the
corresponding parameter window functions.
Note the extremely small uncertainty in the measurements of the
first bands $E1$ and $B1$.

Calculating the expectation value for the nominal concordance model
in each of the 5 bands yields $E$=(0.8,14,13,37,16) and $B$=(0,0,0,0,0) $\mu\,K^2$.
At this point in the ten dimensional parameter space, $\Lambda=5.1$ 
resulting in a PTE of 0.42, and indicating that our data are consistent
with the expected $E$-mode polarization parameterized in this way.
For the no polarization hypothesis, $\Lambda=15.2$ with a ${\rm PTE} = 0.00073$.
While still highly inconsistent with no polarization, this statistic is 
considerably weaker than the equivalent one obtained for the single band analysis 
in \S\ref{sec:EB}, as expected from the increased number of degrees of freedom
in this analysis.  In this ten dimensional space, the probability under the
nopol hypothesis of obtaining a result that is both consistent with the concordance
model and inconsistent with nopol is far lower than that of merely obtaining one that is inconsistent with
nopol.

\subsubsection{Scalar/Tensor}

Predictions exist for the shape of the $E$ and $B$-mode spectra which would result
from primordial tensor perturbations, although their amplitudes
are not well constrained by theory.
In a concordance-type model
such tensor polarization spectra are expected to peak at $l \sim 100$.
Assuming reasonable priors, current measurements of 
the temperature spectrum 
(in which tensor and
scalar contributions will be mixed) suggest
T/S $< 0.2$ \markcite{wang02}({Wang}, {Tegmark}, \& {Zaldarriaga} 2002), where this amplitude ratio
is defined in terms of the tensor and
scalar contributions to the temperature quadrupole $C^T_2$.
We use the distinct polarization angular power spectra for the scalars (our usual
concordance $E$ shape, with $B=0$) and the tensors ($E_T$ and $B_T$) as two components of
a likelihood analysis to constrain the amplitude parameters of these components.
In principle, because the scalar $B$-mode spectrum is zero this approach avoids
the fundamental sample variance limitations arising from using the temperature spectrum
alone.  However, the $E5$/$B5$ analysis (\S\ref{sec:E5B5})
indicates that we have only upper limit to the $E$ or $B-$mode polarization at
the angular scales most relevant ($l \lesssim 200$) for the tensor spectra.  It is therefore
not surprising that our limits on T/S derived from the polarization spectra
as reported in Table \ref{tab:cxpolar_like}, are quite weak.

\begin{table*}
\caption{\label{tab:cxpolar_like}Results of Likelihood Analyses from Polarization Data}
\small%
\begin{center}
\begin{tabular}
[c]{llcrrrrrl}%
		&		&					&		&                                 	& \multicolumn{2}{c}{68\% interval} \\
\rule [-2mm]{0mm}{6mm}
analysis	& parameter 	& $l_{\text{low}}-l_{\text{high}}$	& ML est. 	&$\left(F^{-1}\right)_{ii}^{1/2}$ error	& lower	& upper	& U.L.(95\%)	& units\\
\hline\hline
E/B		& E 		& $-$ 		 			& 0.80  	& $\pm0.28$ 				& 0.56	& 1.10	& $-$		& fraction of concordance E\\
		& B 		& $-$ 		 			& 0.21  	& $\pm0.18$ 				& 0.05	& 0.40	& 0.59		& fraction of concordance E\\
\hline																			 
E/$\beta_E$ 		& E 		& $-$ 		 			& 0.84  	& $\pm0.28$ 				& 0.55	& 1.08	& $-$		& fraction of concordance E\\
		& $\beta_E$ 		& $-$ 		 			& 0.17  	& $\pm1.96$ 				& -1.63	& 1.92	& $-$		& temperature spectral index\\
\hline																			 
E5/B5 		& E1 		& $28-245$ 	        		& -0.50 	& $\pm0.8$ 				& -1.20	& 1.45	& 2.38		& uK$^{2}$\\
		& E2 		& $246-420$ 	 			& 17.1  	& $\pm6.3$ 				&  11.3	& 31.2	& $-$ 		& uK$^{2}$\\
		& E3 		& $421-596$ 	 			& -2.7  	& $\pm5.2$ 				& -10.0	&  4.3	& 24.9		& uK$^{2}$\\
		& E4 		& $597-772$ 	 			& 17.5  	& $\pm16.0$ 				&  3.8	& 40.3	& 47.2		& uK$^{2}$\\
\vspace{1mm}	& E5 		& $773-1050$ 	 			& 11.4  	& $\pm49.0$ 				& -32.5	& 92.3	& 213.2		& uK$^{2}$\\
		& B1 		& $28-245$ 	 			& -0.65 	& $\pm0.65$				& -1.35	& 0.52	& 1.63		& uK$^{2}$\\
		& B2 		& $246-420$ 	 			& 1.3  		& $\pm2.4$ 				& -0.7	& 5.0	& 10.0		& uK$^{2}$\\
		& B3 		& $421-596$ 	 			& 4.8  		& $\pm6.5$ 				& -0.6	& 13.5	& 17.2		& uK$^{2}$\\
		& B4 		& $597-772$ 	 			& 13.0 		& $\pm14.9$ 				& 1.6	& 31.0	& 49.1		& uK$^{2}$\\
		& B5 		& $773-1050$ 	 			& -54.0		& $\pm28.9$ 				& -77.7	& -4.4	& 147.4		& uK$^{2}$\\
\hline																			 
Scalar/Tensor 	& S 		& $-$ 		 			& 0.87		& $\pm0.29$ 				& 0.62	& 1.18	& $-$		& fraction of concordance S\\
		& T 		& $-$ 		 			& -14.3		& $\pm7.5$ 				& -20.4	& -3.9	& 25.4		& T/(S=1)\\
\hline
\end{tabular}
\end{center}
\normalsize
\end{table*}

\subsection{Temperature Data Analyses and Results for $T$ Spectrum}

\subsubsection{T/$\beta_T$}
\label{sec:tspecind}

\begin{figure*}[t]
\begin{center}
\epsfig{file=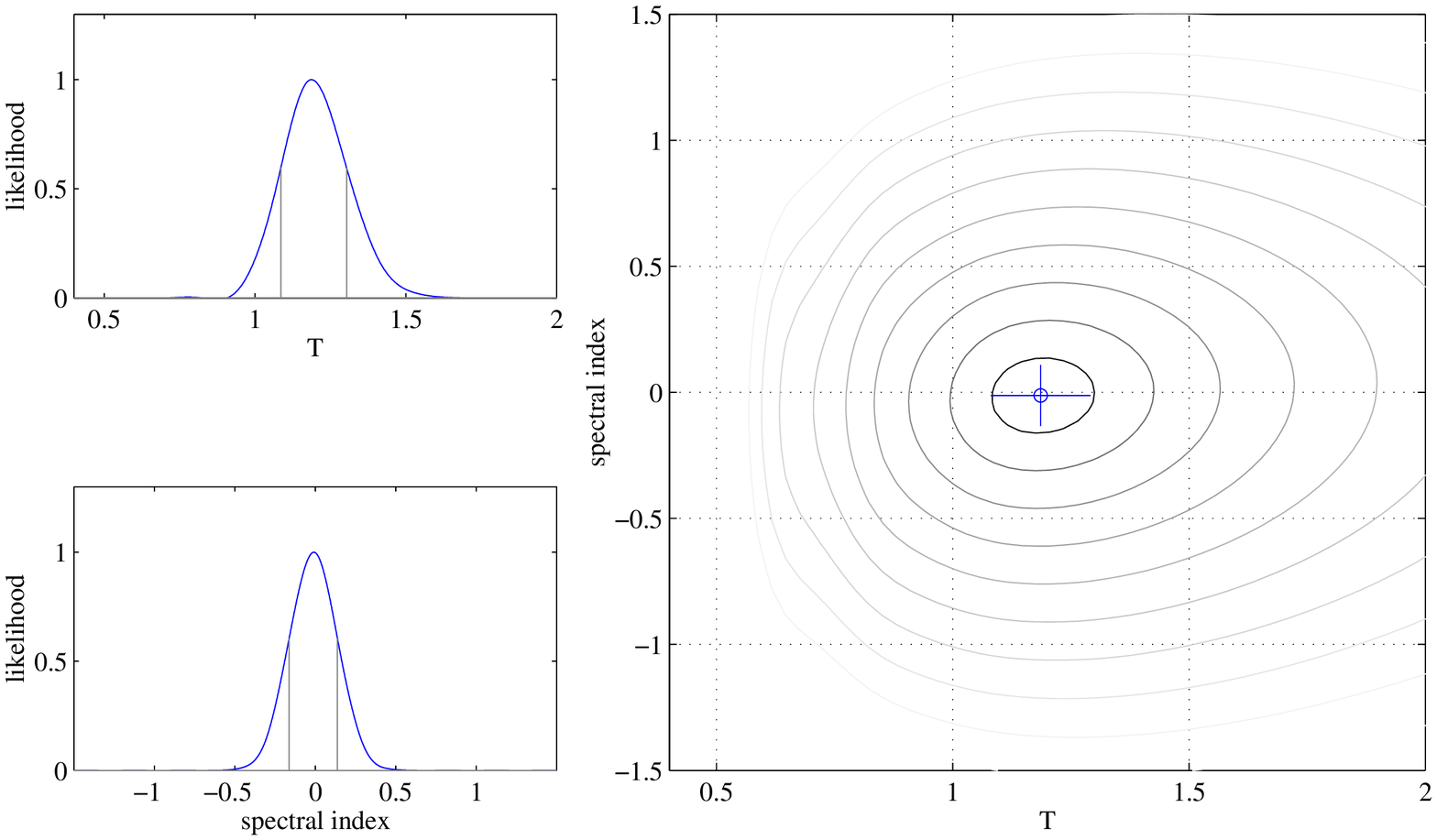,width=3.5in}
\epsfig{file=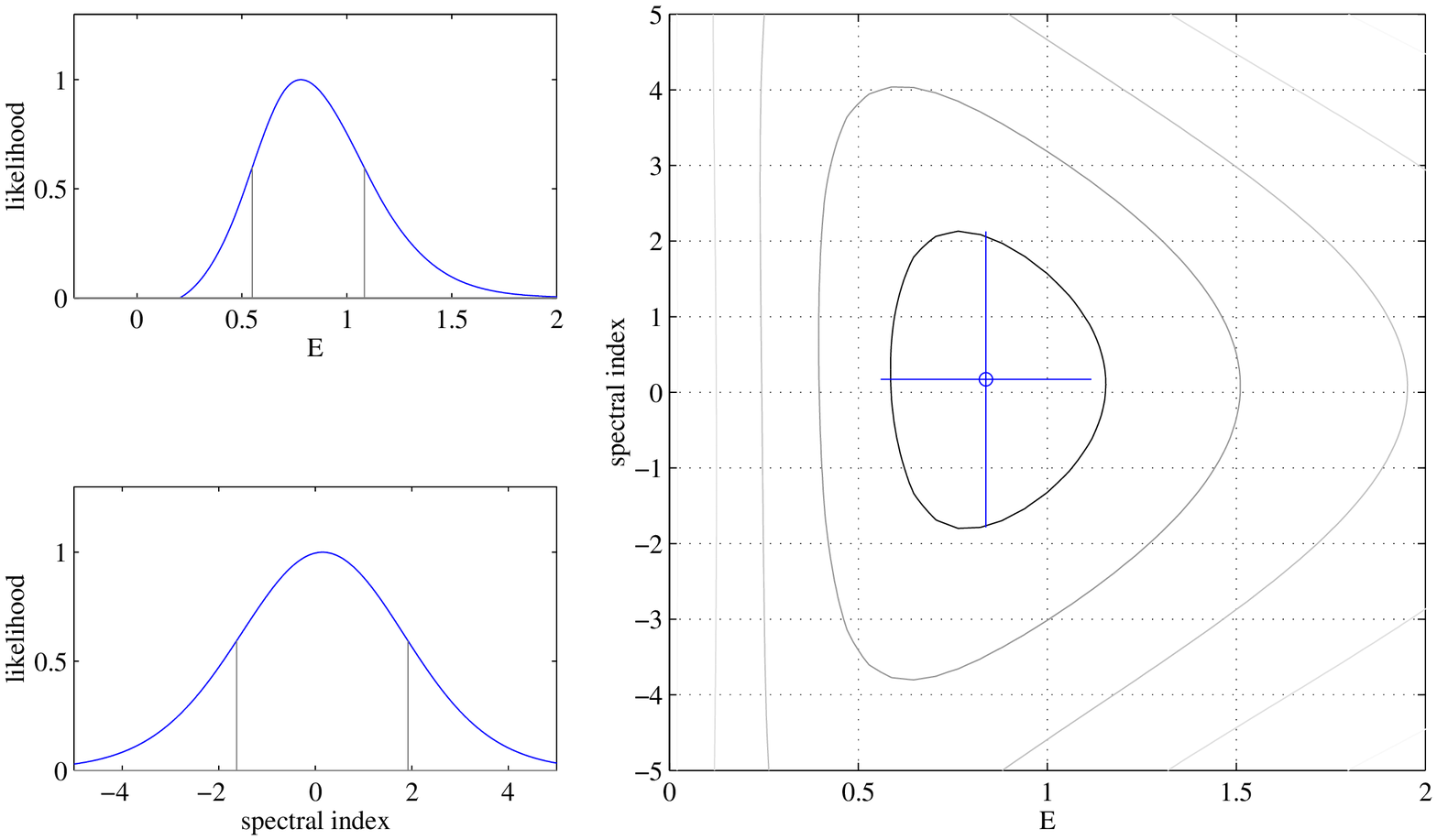,width=3.5in}
\end{center}
\caption{(left) Results from the two parameter shaped bandpower T/$\beta_T$ temperature
analysis assuming the $T$ power spectrum shape as predicted for
the concordance model, and in units relative to that model.
The layout of the plot is analogous to Figure~\ref{fig:EBplot}.
Spectral index is relative to thermal --- in these units synchrotron
emission would be expected to have an index of approximately $-3$.
(right) Results of the similar E/$\beta_E$ analysis performed on the polarization
data.
}
\label{fig:specind}
\end{figure*}

Figure~\ref{fig:specind} shows the results of
a two parameter analysis to determine the amplitude and frequency
spectral index of the temperature signal.
The bandpower shape used is that of the concordance T spectrum,
and the amplitude parameter is expressed in units relative to that spectrum.
The spectral index is relative to the CMB, so that 0 corresponds to a 2.73~K Planck spectrum.
The amplitude of $T$ has a maximum likelihood value of 1.19 (1.09 -- 1.30), and the spectral index $\beta_T = -0.01$ ($-0.16$ -- 0.14). 
While the uncertainty in  the temperature amplitude is dominated by sample variance,
the spectral index is limited only by the sensitivity and fractional
bandwidth of DASI.
Due to the extremely high signal to noise of the temperature data,
the constraints on spectral index are superior to those from 
previous DASI observations (Paper II).

\subsubsection{T5}

The top panel of Figure~\ref{fig:fivebandTEBTE} shows the
results of an analysis using five flat bands to characterize the
temperature spectrum.
These results are completely dominated by the sample variance in the differenced field.
They are consistent with, although less precise than our
previous temperature power spectra described in Paper II; we include them
here primarily to emphasize that DASI makes measurements
simultaneously in all four Stokes parameters and is able to measure temperature
as well as polarization anisotropy.
Note that these results have not been corrected for residual point sources.

\begin{table*}
\caption{\label{tab:copolar_like}Results of Likelihood Analyses from Temperature Data}
\small%
\begin{center}
\begin{tabular}
[c]{llcrrrrl}%
		&		&					&		&                                 	& \multicolumn{2}{c}{68\% interval} \\
\rule [-2mm]{0mm}{6mm}
analysis	& parameter 	& $l_{\text{low}}-l_{\text{high}}$	& ML est. 	&$\left(F^{-1}\right)_{ii}^{1/2}$ error	& lower	& upper	& units\\
\hline\hline																	
T/$\beta_T$ 		& T 		& $-$ 		 			& 1.19  	& $\pm0.11$ 				& 1.09	& 1.30	& fraction of concordance T\\
		& $\beta_T$ 		& $-$ 		 			& -0.01 	& $\pm0.12$ 				& -0.16	& 0.14	& temperature spectral index\\
\hline																		
T5 		& T1 		& $28-245$ 	        		& 6510  	& $\pm1610$				& 5440 	& 9630	& uK$^{2}$\\
		& T2 		& $246-420$ 	 			& 1780  	& $\pm420$ 				& 1480 	& 2490	& uK$^{2}$\\
		& T3 		& $421-596$ 	 			& 2950  	& $\pm540$ 				& 2500 	& 3730	& uK$^{2}$\\
		& T4 		& $597-772$ 	 			& 1910  	& $\pm450$ 				& 1530 	& 2590	& uK$^{2}$\\
		& T5 		& $773-1050$ 	 			& 3810  	& $\pm1210$ 				& 3020 	& 6070	& uK$^{2}$\\
\hline
\end{tabular}
\end{center}
\normalsize
\end{table*}

\subsection{Joint Analyses and Cross Spectra Results: $TE$, $TB$ and $EB$}

\subsubsection{$T,E,TE$}
\label{sec:TETE}

\begin{figure*}[t]
\begin{center}
\epsfig{file=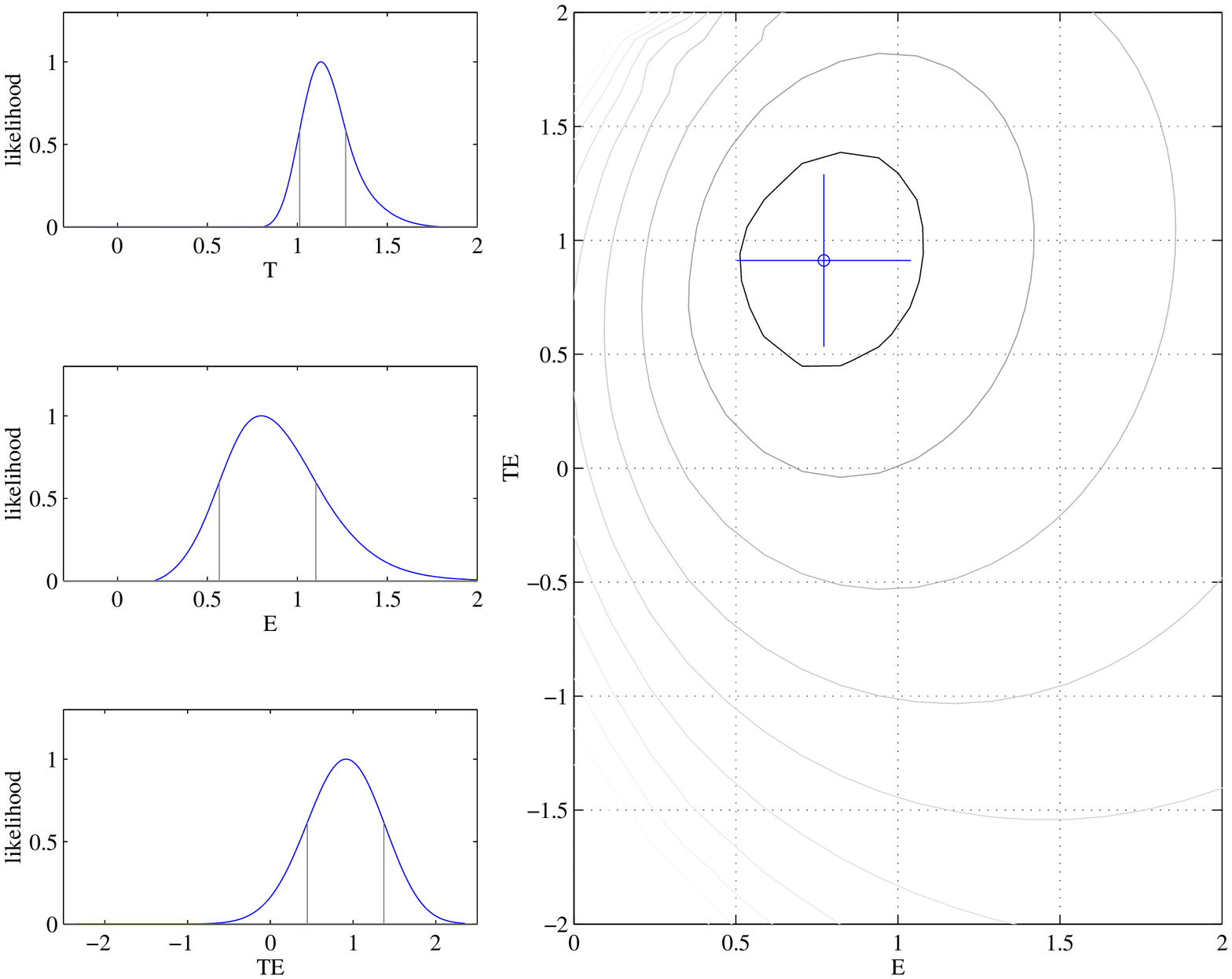,width=5.5in}
\end{center}
\caption{Results from the 3 parameter shaped bandpower $T$/$E$/$TE$ joint analysis,
assuming the spectral shapes as predicted for
the concordance model, and in units relative to that model.
The layout of the plot is analogous to Figure~\ref{fig:EBplot}.
The two dimensional distribution in the right panel is marginalized
over the $T$ dimension.
}
\label{fig:TETEplot}
\end{figure*}

Figure \ref{fig:TETEplot} shows the results of a three parameter single
bandpower analysis of the amplitudes of the $T$ and $E$ spectra, and the
$TE$ cross correlation spectrum.  As before, bandpower shapes based on the
concordance model are used.
The $T$ and $E$ constraints are, as expected, very similar to those for the $E$/$B$, $E$/$\beta_E$ 
and T/$\beta_T$ analyses described above.
The new result here is $TE$ which has a maximum likelihood value of 0.91
with 68\% confidence interval (0.45 to 1.37).
Note that in contrast to the two dimensional likelihoods shown in other figures,
here we see apparent evidence of correlation between the two parameters;
the parameter correlation coefficients from Appendix~\ref{sec:corrtab}
are 0.21 for T/$TE$ and 0.28 for E/$TE$.

Marginalizing over $T$ and $E$, we find that the marginalized likelihood on
TE peaks very near 1, so that $\Lambda(TE=1) = 0.02$ with a PTE of 0.857.
For the no cross correlation hypothesis,
$\Lambda(TE=0) = 1.85$ with an analytic PTE of 0.054 (the PTE
calculated from Monte Carlos is 0.047).
This result represents a detection of the expected $TE$ correlation at
95\% confidence and is particularly interesting in that it suggests
a common origin for the observed temperature and polarization anisotropy.

It has been suggested \markcite{tegmark01}({Tegmark} \& {de Oliveira-Costa} 2001) that an estimator of TE cross correlation
constructed using a $TE=0$ prior may offer greater immunity to systematic errors.
We have confirmed that applying such a technique to our data yields similar results
to the above likelihood analysis, with errors slightly increased as expected.

\subsubsection{$T,E,TE5$}
\label{sec:TETE5}

We have performed a seven parameter analysis using single shaped band
powers for $T$ and $E$, and 5 flat bandpowers for the TE cross correlation;
the $TE$ results from this are shown in the bottom panel of
Figure~\ref{fig:fivebandTEBTE}.
In this analysis the $B$-mode polarization has been explicitly set to zero. 
Again, the $T$ and $E$ constraints are similar to the values for the other
analyses where these parameters appear. 
The $TE$ bandpowers are consistent with the predictions of the 
concordance model.

\subsubsection{$T,E,B,TE,TB,EB$}

Finally, we describe the results of a six shaped bandpower analysis
for the three individual spectra $T$, $E$ and $B$, together with the three
possible cross correlation spectra $TE$, $TB$ and $EB$.  We include the $B$
cross-spectra for completeness, though there is little evidence for
any $B$-mode signal.
Because there are no predictions for the shapes of the $TB$ or $EB$
spectra, we preserve the symmetry of the analysis between $E$ and $B$ by
simply parameterizing them in terms of the $TE$ and $E$ spectral shapes.  The
results for $T$, $E$, $B$ and $TE$ are similar to as before, with no detection
of $EB$ or $TB$.

\begin{table*}
\caption{\label{tab:joint_like}Results of likelihood analyses from joint temperature-polarization dataset}
\small%
\begin{center}
\begin{tabular}
[c]{llcrrrrl}%
		&		&					&		&                                 	& \multicolumn{2}{c}{68\% interval} \\
\rule [-2mm]{0mm}{6mm}
analysis	& parameter 	& $l_{\text{low}}-l_{\text{high}}$	& ML est. 	&$\left(F^{-1}\right)_{ii}^{1/2}$ error	& lower	& upper	& units\\
\hline\hline																	
T/E/TE		& T 		& $-$ 		 			& 1.13  	& $\pm0.10$ 				& 1.05	& 1.29	& fraction of concordance T\\
		& E 		& $-$ 		 			& 0.77  	& $\pm0.27$ 				& 0.57	& 1.10	& fraction of concordance E\\
		& TE		& $-$ 		 			& 0.91  	& $\pm0.38$ 				& 0.45	& 1.37	& fraction of concordance TE\\
\hline																		
T/E/TE5		& T 		& $-$ 		 			& 1.12  	& $\pm0.10$ 				& 1.09	& 1.31	& fraction of concordance T\\
\vspace{1mm}	& E  		& $-$ 		 			& 0.81  	& $\pm0.28$ 				& 0.71 	& 1.36	& fraction of concordance E\\
      		& TE1 		& $28-245$ 	        		& -24.8 	& $\pm32.2$ 				& -55.3	& 24.7	& uK$^{2}$\\
		& TE2 		& $246-420$ 	 			& 92.3  	& $\pm38.4$ 				& 44.9 	& 151.1	& uK$^{2}$\\
		& TE3 		& $421-596$ 	 			& -10.5 	& $\pm48.2$ 				& -60.1	& 52.0	& uK$^{2}$\\
		& TE4 		& $597-772$ 	 			& -66.7 	& $\pm74.3$ 				&-164.6	& 9.5 	& uK$^{2}$\\
            	& TE5 		& $773-1050$ 	 			& 20.0  	& $\pm167.9$ 				&-130.3	& 172.3	& uK$^{2}$\\
\hline																		
T/E/B/TE/TB/EB	& T 		& $-$ 		 			& 1.13		& $\pm0.10$ 				& 1.03	& 1.27	& fraction of concordance T\\
              	& E 		& $-$ 		 			& 0.75		& $\pm0.26$ 				& 0.59	& 1.19	& fraction of concordance E\\
              	& B 		& $-$ 		 			& 0.20		& $\pm0.18$ 				& 0.11	& 0.52	& fraction of concordance E\\
              	& TE		& $-$ 		 			& 1.02		& $\pm0.37$ 				& 0.53	& 1.49	& fraction of concordance TE\\
              	& TB		& $-$ 		 			& 0.53		& $\pm0.32$ 				& 0.08	& 0.82	& fraction of concordance TE\\
              	& EB		& $-$ 		 			& -0.16		& $\pm0.16$ 				&-0.38	& 0.01	& fraction of concordance E\\
\hline
\end{tabular}
\end{center}
\normalsize
\end{table*}

\section{Systematic Uncertainties}

\label{sec:systematics}

\subsection{Noise, Calibration, Offsets and Pointing}

In this section, we discuss the effect of systematic uncertainties on
the likelihood results. We have repeated each of the nine analyses
with alternative assumptions about the various effects which we have
identified, and for which there is a well-defined uncertainty.  The
results of these tests are described below.

Much of the effort of the data analysis presented in this paper has
gone into investigating the consistency of the data with the noise
model discussed in \S\ref{sec:noisemodel}.  As discussed in that
section, we find no discrepancies between complementary noise
estimates on different timescales, to a level $\ll1\%$.  As discussed
in \S\ref{sec:chi2tests}, numerous consistency tests on subsets of the
co-polar and cross-polar visibility data show no evidence for
an error 
in the noise scaling to a similar level.  When we
re-evaluate each of the analyses described in \S\ref{sec:lhresults}
with the noise scaled by $1\%$, the shift in the maximum likelihood
values for all parameters is entirely negligible.

In \S\ref{sec:noisemodel}, we reported the presence of detectable
correlations between real/imaginary visibilities and between
visibilities from different bands of the same baseline.
When these correlations
are added to the covariance matrix at the measured level, either
separately or together, the effects are again negligible; the largest
shift is in the highest-$l$ bin of the $E$ spectrum from the 
$E5/B5$ analysis
(\S\ref{sec:E5B5}), where the
power shifts by $\sim2~\mu{\rm K}^2$.

The absolute cross-polar phase offsets, if uncorrected, will mix power between
$E$ and $B$; these phases have been independently determined from
wire-grid calibrations and observations of the Moon, and found to
agree to within the measurement uncertainties of $\sim0\fdg4$ (Paper IV).  
Reanalysis of the
data with the measured phase offsets shifted by $2\deg$ demonstrates
that the likelihood results are immune to errors of this magnitude;
the largest effect occurs in the highest-$l$ bin of the $TE$ spectrum 
from the $T,E,TE5$
analysis (\S\ref{sec:TETE5}), where the power shifts by $\sim30~\mu{\rm K}^2$.

The on-axis leakages described in \S\ref{sec:onaxisleak} will mix
power from $T$ into $E$ and $B$, and the data are corrected for this
effect before input to the likelihood analyses.  When these analyses
are performed without the leakage correction, the largest effects
appear in the shaped $TE$ amplitude analysis (\S\ref{sec:TETE}), and the lowest-$l$ bin of $TE5$ from the
$T,E,TE5$ analysis (\S\ref{sec:TETE5}); all shifts are tiny compared to the 
68\%
confidence intervals.  As the leakage correction itself has little
impact on the results, the uncertainties in the correction which are
at the $<1\%$ level will have no noticeable effect.

As described in \S\ref{sec:offaxisleak}, the off-axis leakage from the
feeds is a more significant effect, and is accounted for in the
likelihood analysis by modeling its contribution to the covariance
matrix.  When this correction is not applied, the $E,B$ results
(\S\ref{sec:EB}) shift
by $\sim4\%$ and $\sim2\%$, respectively, as
expected from simulations of this effect. 
Although this bias is
already a small effect, the simulations show that the correction 
removes it completely to the degree we understand the
off-axis leakage. Uncertainties in the 
leakage profiles of the order of the fit residuals (see Paper IV)
lead to a bias of less than 1\%.

The pointing accuracy of the telescope is
measured to be better than $2'$
and the rms tracking errors are
$<20''$, as discussed in Papers I and II, this is more than sufficient 
for the characterization of
CMB anisotropy at the much larger angular scales measured by
DASI.

Absolute calibration of the telescope was achieved through
measurements of external thermal loads, transferred 
to the calibrator RCW38. The dominant uncertainty
in the calibration is due to temperature and coupling
of the thermal loads. As discussed in Paper II, 
we estimate an overall
calibration uncertainty of 8\% (1\,$\sigma$), expressed as a
fractional uncertainty on the $C_l$ bandpowers (4\% in $\Delta T/T$).
This applies equally to the temperature and polarization data
presented here.

\subsection{Foregrounds}

\subsubsection{Point sources}

\label{sec:pointsources}

The highest sensitivity point source catalog in our observing
region is the 5~GHz PMN survey \markcite{wright94}({Wright} {et~al.} 1994).
For our first season temperature analysis described in Papers
I and II we projected out known sources using this catalog.
We have kept to this procedure for the temperature data
presented here, projecting the same set of sources as before.

Unfortunately the PMN survey is not polarization sensitive.
We note that the distribution of point source polarization fractions
is approximately exponential (see below). 
Total intensity is thus a poor indicator of polarized
intensity and it is therefore not sensible to project out
the PMN sources in our polarization analysis.

Our polarization fields were selected for the absence of any
significant point source detections in the first season data.
No significant detections are found in the 2001 -- 2002 data, either in the
temperature data, which are dominated by CMB anisotropy,  or in the 
polarization data.

To calculate the expected contribution of un-detected point sources
to our polarization results we would like to know the distribution
of polarized flux densities, but unfortunately no such information exists
in our frequency range.
However, to make an estimate, we use the distribution of total
intensities, and then assume a distribution of polarization fractions.
We know the former distribution quite well from our own first season
32-field data where we detect 31 point sources and determine that
$dN/dS_{31}=(32\pm7) S^{(-2.15\pm0.20)}$~Jy$^{-1}$~Sr$^{-1}$
in the range 0.1 to 10~Jy.
This is consistent, when extrapolated to lower flux densities,
with a result from the CBI experiment valid in the range 5--50~mJy~\markcite{mason02}({Mason} {et~al.} 2002).
The distribution of point source polarization fractions at 5 GHz
can be characterized by an exponential with a mean of 3.8\% \markcite{zukowski99}({Zukowski} {et~al.} 1999);
data of somewhat lower quality at 15~GHz are consistent with the
same distribution \markcite{simard81b}({Simard-Normandin},  {Kronberg}, \& {Neidhoefer} 1981b).
Qualitatively one expects the polarization fraction of
synchrotron-dominated sources to initially rise with frequency,
and then plateau or fall, with the break point at frequencies
$\ll5$~GHz (see \markcite{simard81a}{Simard-Normandin},  {Kronberg}, \& {Button} (1981a) for an example).
In the absence of better data we have conservatively assumed the exponential distribution
mentioned above continues to hold at 30~GHz. 

We proceed to estimate the effect of point sources by Monte Carlo
simulation, generating realizations using the total intensity and
polarization fraction distributions mentioned above.
For each realization, we generate simulated DASI data by
adding a realization of CMB anisotropy and appropriate instrument noise.
The simulated data are tested for evidence of point sources and those 
realizations that show statistics similar to the real data 
are kept.
The effect of off-axis leakage, which is described and quantified in
paper IV, is included in these calculations.

When the simulated data are passed through the $E/B$ analysis
described in \S\ref{sec:eb_anal}, the mean bias of the $E$ parameter
is 0.04 with a standard deviation of 0.05; in 95\% of
cases the shift distance in the $E/B$ plane is less than
0.13.
We conclude that the presence of point sources consistent our observed data 
have a relatively small effect on our polarization results.

\subsubsection{Diffuse Foregrounds}
\label{sec:diffusefore}

In Paper I, we gave estimates of the total intensity of synchrotron,
free-free and thermal dust emission in the region of our fields,
showing that the expected amplitudes are very small.  This was
confirmed in Paper II by a template based cross correlation analysis
which showed that the contribution of each of these foregrounds to our
temperature anisotropy results were negligible.

The expected fractional polarization of the CMB is of order 10\%.
The corresponding number for free-free emission is less than
1\% and thermal dust emission may be polarized by
several percent \markcite{hildebrand00}(see, e.g., {Hildebrand} {et~al.} 2000).
Therefore if free-free and dust emission did not contribute
significantly to our temperature anisotropy results they are not
expected to contribute to the polarization.
Synchrotron emission on the other hand can in principle be up to 70\%
polarized, and is by far the greatest concern; what was a negligible
contribution in the temperature case could be a significant one in
polarization.

There are no published polarization maps in the region of our fields.
Previous attempts to estimate the angular power spectrum of
polarized synchrotron emission have been guided by surveys of the
Galactic plane at frequencies of 2 -- 3~GHz \markcite{tegmark_ehd99}(Tegmark {et~al.} 2000).
These maps show much more small scale structure in polarization than
in temperature, but this is mostly induced by Faraday rotation, an
effect which is negligible at 30~GHz.
Additionally, since synchrotron emission is highly concentrated
in the disk of the Galaxy it is not valid to assume
that the angular power spectrum at low Galactic latitudes has much
to tell us about that at high.

Our fields lie at Galactic latitude $-58\fdg4$ and $-61\fdg9$.
The brightness of the IRAS 100 micron and Haslam 408~MHz \markcite{haslam81}({Haslam} {et~al.} 1981)
maps within our fields lie at the $6\%$ and $25\%$ points, respectively, of
the integral distributions taken over the whole sky.
There are several strong pieces of evidence from the DASI dataset itself
that the polarization results 
described in this paper are free of significant synchrotron contamination.
The significant $TE$ correlation shown in Figure~\ref{fig:TETEplot} indicates
that the temperature and $E$-mode signal have a common origin. 
The tight constraints on the temperature anisotropy spectral index
require that this common origin has a spectrum consistent with CMB. 
Galactic synchrotron emission is known to have a temperature spectral index
of $-2.8$ \markcite{platania98}({Platania} {et~al.} 1998), with evidence for steepening
to $-3.0$ at frequencies above 1 -- 2~GHz \markcite{banday91}({Banday} \& {Wolfendale} 1991).
At frequencies where Faraday depolarization is negligible ($>10$~GHz),
the same index will also apply for polarization.
The dramatically tight constraint on the temperature spectral index
of 0.01 ($-0.16$ -- 0.14) indicates that any component of the temperature signal 
coming from synchrotron emission is negligibly small in comparison to the CMB.
More significantly, the constraint on the $E$-mode spectral index 
$\beta_E =0.17$ ($-1.63$ -- 1.92) disfavors synchrotron at nearly $2\sigma$.
A third, albeit weaker, line of argument is that 
a complex synchrotron emitting structure is not
expected to
produce a projected brightness distribution 
which prefers $E$-mode polarization over $B$-mode.
Therefore, the result in
Figure~\ref{fig:EBplot} could be taken as further evidence that the
signal we are seeing is not due to synchrotron emission.

\section{Conclusion}
\label{sec:conclusion}

In this paper, we present the first detection of polarization of 
the CMB.  These results are the product of two years of
observations with the DASI telescope within two $3\fdg4$ FWHM fields.  For
the observations described here, DASI was reconfigured with
achromatic polarizers to provide sensitivity in all four Stokes
parameters.  As described in Paper IV, observations of both polarized and unpolarized
astronomical sources give us confidence that the gain and instrumental
polarization of the telescope have been precisely characterized. 

We have performed extensive consistency tests on various splits and subsets of
the visibility data.  For those modes expected to
have high s/n, 
a simple comparison of the polarization data with the measured instrumental 
noise results in a
robust detection of a polarized signal with a significance of
approximately 5$\sigma$.  
These tests show no indication of systematic
contamination and strongly support a celestial origin
of the polarized
signal.
We employ a full likelihood analysis to determine confidence intervals for 
temperature and polarization models parameterized by shaped and flat band 
powers.
Unlike the DASI temperature angular power spectrum reported in Paper II, the 
temperature power spectrum presented in this paper is strongly dominated by 
sample variance.
However, the high s/n achieved in the deep polarization presented here 
permits a precise 
determination of the spectral index of the CMB temperature anisotropy,
$\beta_T = -0.01$ ($-0.16$ -- 0.14 at 68\% confidence).

A likelihood ratio test is used to demonstrate the agreement of the   
observed CMB temperature and polarization anisotropy signals with a 
concordance $\Lambda$CDM model,  and strongly rejects models
without CMB polarization. 
From this analysis we determine that we have detected
$E$-mode CMB polarization with a significance of $4.9\sigma$. 
Specifically, assuming a shape for the
power spectrum consistent
with previous temperature measurements, the level found for the
$E$-mode polarization is 0.80 (0.56 -- 1.10), where the 
predicted level given previous temperature data is 0.9 -- 1.1.

The spectral index determined for the observed $E$-mode polarization signal,
$\beta_E = 0.17$ ($-1.63$ -- 1.92), is 
consistent with CMB. At 95\% confidence, an upper limit of 0.59 is set to the level of $B$-mode polarization with the
same shape and normalization as the $E$-mode spectrum.  The $TE$
correlation of the temperature and $E$-mode polarization is detected at 
95\% confidence, and also found to be consistent with
predictions.

We have considered the possibility that our results are contaminated by foreground 
emission in the form of a distribution of polarized
radio point sources and high Galactic latitude synchrotron emission. 
Simulated distributions of radio sources are shown to contribute insignificant
polarization compared to the observed signal.
The strongest constraints against diffuse synchrotron emission come from
the DASI dataset itself.
The observed $TE$ correlation, combined with the precisely thermal
spectrum of the temperature anisotropy creates a compelling argument that 
the $E$-mode polarization we observe was created at the surface of last 
scattering.  
Although the constraint on the $E$-mode polarization spectral index is not nearly 
as strong as those for the temperature anisotropy, this result is 
incompatible with
Galactic synchrotron as the source of the observed polarization at 
nearly $2\sigma$.  In general, foregrounds are expected to produce
comparable amplitude in both $E$- and $B$-mode spectra. Our data
therefore provide additional evidence against a strong
contribution from foreground emission 
to the degree
that our results limit the ratio of $B$- to $E$-mode
polarization.

The likelihood results and tests to which we have subjected the data
provide self-consistent and strong support for the detection of
the polarization induced on the CMB at the surface of last scattering.
These results provide strong validation of the underlying
theoretical framework for the origin of CMB anisotropy and lend
confidence to the values of the cosmological parameters that have been
derived from CMB measurements.

\acknowledgments

We are grateful for the competent and dedicated efforts of
Ben Reddall and Eric Sandberg, who wintered over at the National
Science Foundation (NSF)
Amundsen-Scott South Pole research station, to keep DASI running smoothly.  We are
indebted to Mark Dragovan for his role in making DASI a reality,
and to the Caltech CBI team led by Tony Readhead, in particular, to
Steve Padin, John Cartwright, Martin Shepherd, and John Yamasaki for the development 
of key hardware and software.

We are indebted to the Center for Astrophysical Research in Antarctica
(CARA), in particular to the CARA polar operations staff.
We are grateful for valuable contributions from Kim Coble, Allan Day,
Gene Drag, Jacob Kooi, Ellen LaRue, Mike Loh, Bob Lowenstein, Stephan
Meyer, Nancy Odalen, Bob, Dave and Ed Pernic, Bob Spotz and Mike Whitehead.
We thank Raytheon Polar Services for their support of the DASI project.  
We have benefited from many interactions with the Center for
Cosmological Physics members and visitors.  In particular, we
gratefully acknowledge many illuminating conversations with Wayne Hu
on the intricacies of CMB polarization and valuable suggestions from
Steve Meyer, Mike Turner and Bruce Winstein on the presentation of
these results, and we thank Lloyd Knox and Arthur Kosowsky for
bringing the Markov technique to our attention.
We thank the observatory staff of the Australia Telescope Compact
Array, in particular Bob Sault and Ravi Subrahmanyan, for their
generosity in providing point source observations of the DASI fields.

This research was initially supported by the NSF
under a cooperative agreement (OPP 89-20223) with CARA, a
NSF Science and Technology Center. It is
currently supported by NSF grant OPP-0094541. JEC gratefully
acknowledges support from the James S. McDonnell Foundation and the
David and Lucile Packard Foundation.  JEC and CP gratefully
acknowledge support from the Center for Cosmological Physics.

\appendix

\section{Likelihood Correlation Matrices}
\label{sec:corrtab}

Below we tabulate the correlation matrices for our various
likelihood analyses to allow the reader to gauge the degree
to which each parameter has been determined independently.
The covariance matrix is the inverse of the Fisher matrix
and the correlation matrix is defined as the covariance matrix
normalized such that the diagonal is unity, i.e.,
$C=F^{-1}$ and $R_{ij} = C_{ij}/\sqrt{C_{ii}C_{jj}}$.

\subsection{Correlation Coefficient Matrices for 2 Parameter Analyses}
\small
\begin{center}
\begin{tabular}{cccccccccccccc}
        $\phm{-}{\rm E}$  &         $\phm{-}{\rm B}$  && $\phm{-}{\rm E}$ & $\phm{-}\beta_E$ && $\phm{-}{\rm S}$ & $\phm{-}{\rm T}$ && \phm{-}T & $\phm{-}\beta_T$ \\ 
\cline{1-2} \cline{4-5} \cline{7-8} \cline{10-11}
$\phm{-}1$&  $-0.046$&&$\phm{-}1$&  $-0.082$ &&$\phm{-}1$&  $-0.339$&&$\phm{-}1$& $0.023$\\
          &$\phm{-}1$&&          &$\phm{-}1$ &&          &$\phm{-}1$&&          &     $1$\\
\cline{1-2} \cline{4-5} \cline{7-8} \cline{10-11}
\end{tabular} 
\end{center}
\normalsize

\subsection{Correlation Coefficient Matrices for $E5$/$B5$ Analysis}
\small
\begin{center}
\begin{tabular}{cccccccccc}
$\phm{-}{\rm E1}$&$\phm{-}{\rm E2}$&$\phm{-}{\rm E3}$&$\phm{-}{\rm E4}$&$\phm{-}{\rm E5}$&$\phm{-}{\rm B1}$&$\phm{-}{\rm B2}$&$\phm{-}{\rm B3}$&$\phm{-}{\rm B4}$  &        $\phm{-}{\rm B5}$  \\ 
\hline 
     $1$&    $-0.137$&$\phm{-}0.016$&      $-0.002$&$\phm{-}0.000$&      $-0.255$&$\phm{-}0.047$&      $-0.004$&$\phm{-}0.000$&$\phm{-}0.000$ \\ 
        &  $\phm{-}1$&      $-0.117$&$\phm{-}0.014$&      $-0.002$&$\phm{-}0.024$&      $-0.078$&$\phm{-}0.004$&$\phm{-}0.000$&$\phm{-}0.000$ \\ 
        &            &    $\phm{-}1$&      $-0.122$&$\phm{-}0.015$&      $-0.003$&$\phm{-}0.010$&      $-0.027$&$\phm{-}0.003$&      $-0.001$ \\ 
        &            &              &    $\phm{-}1$&      $-0.119$&$\phm{-}0.000$&      $-0.001$&$\phm{-}0.002$&      $-0.016$&$\phm{-}0.003$ \\ 
        &            &              &              &    $\phm{-}1$&$\phm{-}0.000$&$\phm{-}0.000$&$\phm{-}0.000$&$\phm{-}0.002$&      $-0.014$ \\ 
        &            &              &              &              &    $\phm{-}1$&      $-0.226$&$\phm{-}0.022$&      $-0.002$&$\phm{-}0.000$ \\ 
        &            &              &              &              &              &    $\phm{-}1$&      $-0.097$&$\phm{-}0.011$&      $-0.002$ \\ 
        &            &              &              &              &              &              &    $\phm{-}1$&      $-0.111$&$\phm{-}0.018$ \\ 
        &            &              &              &              &              &              &              &    $\phm{-}1$&      $-0.164$ \\ 
        &            &              &              &              &              &              &              &              &           $1$ \\ 
\hline 
\end{tabular} 
\end{center}
\normalsize

\subsection{Correlation Coefficient Matrix for $T5$ Analysis}
\small
\begin{center}
\begin{tabular}{ccccc}
       $\phm{-}{\rm T1}$&$\phm{-}{\rm T2}$&$\phm{-}{\rm T3}$&$\phm{-}{\rm T4}$&$\phm{-}{\rm T5}$ \\ 
\hline 
 $\phm{-}1$&  $-0.101$&$\phm{-}0.004$&   $-0.004$&    $-0.001$  \\ 
           &$\phm{-}1$&      $-0.092$&   $-0.013$&    $-0.011$  \\ 
           &          &    $\phm{-}1$&   $-0.115$&    $-0.010$  \\ 
           &          &              & $\phm{-}1$&    $-0.147$  \\ 
           &          &              &           &  $\phm{-}1$  \\ 
\hline 
\end{tabular} 
\end{center}
\normalsize

\subsection{Correlation Coefficient Matrix for $T$/$E$/$TE$ Analysis}
\small
\begin{center}
\begin{tabular}{ccc}
$\phm{-}{\rm T}$&$\phm{-}{\rm E}$&$\phm{-}{\rm TE}$  \\ 
\hline 
 $\phm{-}1$&$\phm{-}0.017$&$\phm{-}0.207$  \\ 
           &    $\phm{-}1$&$\phm{-}0.282$  \\ 
           &              &    $\phm{-}1$  \\ 
\hline 
\end{tabular} 
\end{center}
\normalsize

\subsection{Correlation Coefficient Matrix for $T$/$E$/$TE5$ Analysis}
\small
\begin{center}
\begin{tabular}{ccccccc}
$\phm{-}{\rm T}$&$\phm{-}{\rm E}$&$\phm{-}{\rm TE1}$&$\phm{-}{\rm TE2}$&$\phm{-}{\rm TE3}$&$\phm{-}{\rm TE4}$&$\phm{-}{\rm TE5}$ \\ 
\hline 
$\phm{-}1$&$\phm{-}0.026$&  $-0.071$&$\phm{-}0.202$&      $-0.018$&      $-0.075$&$\phm{-}0.008$  \\ 
          &    $\phm{-}1$&  $-0.067$&$\phm{-}0.339$&      $-0.023$&      $-0.090$&$\phm{-}0.008$  \\ 
          &              &$\phm{-}1$&      $-0.076$&$\phm{-}0.006$&$\phm{-}0.011$&      $-0.001$  \\ 
          &              &          &    $\phm{-}1$&      $-0.078$&      $-0.039$&$\phm{-}0.004$  \\ 
          &              &          &              &    $\phm{-}1$&      $-0.056$&$\phm{-}0.004$  \\ 
          &              &          &              &              &    $\phm{-}1$&      $-0.066$  \\ 
          &              &          &              &              &              &    $\phm{-}1$  \\ 
\hline 
\end{tabular} 
\end{center}
\normalsize

\subsection{Correlation Coefficient Matrix for $T$/$E$/$B$/$TE$/$TB$/$EB$ Analysis}
\small
\begin{center}
\begin{tabular}{cccccc}
$\phm{-}{\rm T}$&$\phm{-}{\rm E}$&$\phm{-}{\rm B}$&$\phm{-}{\rm TE}$&$\phm{-}{\rm TB}$&$\phm{-}{\rm EB}$  \\ 
\hline 
$\phm{-}1$&$\phm{-}0.026$&$\phm{-}0.004$&$\phm{-}0.230$&$\phm{-}0.136$&$\phm{-}0.033$  \\ 
          &    $\phm{-}1$&      $-0.027$&$\phm{-}0.320$&      $-0.040$&      $-0.182$  \\ 
          &              &    $\phm{-}1$&      $-0.027$&$\phm{-}0.219$&      $-0.190$  \\ 
          &              &              &    $\phm{-}1$&      $-0.150$&$\phm{-}0.109$  \\ 
          &              &              &              &    $\phm{-}1$&$\phm{-}0.213$  \\ 
          &              &              &              &              &    $\phm{-}1$  \\ 
\hline 
\end{tabular} 
\end{center}
\normalsize

\bibliography{}

\end{document}